\documentclass[12pt]{article}
\usepackage[utf8]{inputenc}
\usepackage{amsmath,setspace,geometry}
\usepackage{amsthm}
\usepackage{amsfonts}
\usepackage[shortlabels]{enumitem}
\usepackage{rotating}
\usepackage{pdflscape}
\usepackage{graphicx}
\usepackage{bbm}
\usepackage{comment}
\usepackage[dvipsnames]{xcolor}
\usepackage{hyperref}
\hypersetup{colorlinks=true, linkcolor= BrickRed, citecolor = BrickRed, filecolor = BrickRed, urlcolor = BrickRed, hypertexnames = true}
\usepackage[longnamesfirst]{natbib}

\bibpunct[:]{(}{)}{,}{a}{}{,}
\usepackage[english]{babel}
\usepackage{float}
\usepackage{caption}
\usepackage{subcaption}
\usepackage{booktabs}
\usepackage{pdfpages}
\usepackage{adjustbox}
\usepackage{tabularx}
\usepackage{threeparttable}
\usepackage{lscape}
\usepackage{bm}
\usepackage{dsfont}
\usepackage{mathtools}
\mathtoolsset{showonlyrefs=true}

\usepackage{multirow}
\usepackage{booktabs}
\usepackage{adjustbox}
\bibpunct[:]{(}{)}{,}{a}{}{,}
\newenvironment{minilinespace}{
\baselineskip = 4mm
}

\usepackage{setspace}
\begin{document}
\title{{Are Final Market Prices Sufficient for Information Aggregation? Evidence from Last-Minute Dynamics in Parimutuel Betting}\thanks{The authors are grateful to Pierre-André Chiappori, Nobuyuki Hanaki, Yuichiro Kamada, Hayato Kanayama, Michihiro Kandori, Shunya Noda, Yasuhiro Shirata, Ryoji Yanashima, and Takahiro Watanabe for their insightful comments and suggestions. We also thank the participants in the SWET 2025 Market Design session and the 19th Annual Meeting of the Association of Behavioral Economics for helpful feedback and discussions. An unrefereed extended abstract of this paper appeared in the proceedings of the 19th Annual Meeting, published in the \textit{Journal of Behavioral Economics and Finance}, Vol. 18, Special Issue (2025), S1--S4. This research uses data from the JRA-VAN database (a licensed service providing structured JRA data), with official permission granted by JRA-VAN for academic use. Financial support from JST ERATO Grant Number JPMJER2301 is gratefully acknowledged.}}
\author{Hiroaki Hanyu\thanks{\href{mailto:}{bd251004@g.hit-u.ac.jp}, Hitotsubashi University}\and Shunsuke Ishii\thanks{\href{mailto:}{s.ishii@lse.ac.uk}, London School of Economics and Political Science} \and  Suguru Otani\thanks{\href{mailto:}{suguru.otani@e.u-tokyo.ac.jp}, Market Design Center, Graduate School of Economics, University of Tokyo}\and  Kazuhiro Teramoto\thanks{\href{mailto:}{k.teramoto@r.hit-u.ac.jp}, Graduate School of Economics, Hitotsubashi University.}}

 \date{
First version: September 18, 2025\\
Current version: July 7, 2026 
}

\maketitle

\begin{abstract}

This study presents evidence challenging the practice of inferring risk preferences, probability perceptions, and beliefs from parimutuel betting markets, where wagers cannot be made contingent on final odds. Using interim odds from horse racing, we show that expected returns depend not only on final odds but also on the path through which they are reached: horses with final-five-minute odds declines earn higher realized returns than horses with similar final odds. We rationalize this path dependence using a two-period extension of the information-based model \`{a} la Ottaviani--S{\o}rensen, in which final-stage odds declines arise from informed bettors' late wagers based on private signals. The model also highlights that final odds--return patterns alone may not distinguish information aggregation from probability distortions.

\end{abstract}

\newpage

\section{Introduction}

Market prices under uncertainty reflect market participants' beliefs, preferences, and information. Betting markets offer a useful setting for studying how these underlying objects can be inferred from field data, because both market prices---in the form of odds---and outcomes are observed. Indeed, a large literature has used final odds--return patterns in horse-race betting to infer risk preferences, probability perceptions, or heterogeneous beliefs, often in frameworks analogous to asset-pricing models \citep[e.g.,][]{Weitzman, Ali, jullien2000estimating, snowberg2010explaining, gandhi2015does, chiappori2019aggregate}.\footnote{For a broader review of risk-preference estimation using field data, see \citet{barseghyan2018estimating}; for experimental and theoretical perspectives, see \citet{plott2003parimutuel}, \citet{koessler2012information}, \citet{kajii2017favorite}, and \citet{gillen2017pari}.}

Inferring preferences or beliefs from final odds--return patterns, however, requires a view of how market prices---odds in this context---are formed. The parimutuel system, a common institutional format in horse-race betting, is not a standard market-clearing price mechanism. Bettors do not trade at prices that clear supply and demand, nor can they submit limit orders or otherwise make their wagers contingent on final odds. Instead, interim and final odds are mechanically determined by aggregate betting shares, with final odds realized only after the betting window closes. This institutional feature implies that final odds--return patterns may reflect not only preferences or beliefs, but also the process through which private information is incorporated into odds. \citet{ottaviani2009surprised} formalize this idea by showing that the favorite--longshot bias (FLB)---longshots earn lower average returns than favorites---can arise \emph{ex post} from information aggregation, even under risk neutrality.\footnote{For surveys of the FLB and betting-market efficiency, see \citet{thaler1988anomalies}, \citet{hausch1995efficiency}, \citet{ottaviani2008favorite}, and \citet{jullien2008empirical}.}

This paper builds on this information-based perspective by focusing on the evolution of odds before they become final. Motivated by the incentive for informed bettors to delay wagering \citep{ottaviani2006timing}, we ask whether odds movements near market closure contain information about subsequent returns. We use interim odds to trace how market shares evolve before final odds are realized, and study whether realized returns are related not only to final odds but also to the trajectory through which those odds are reached.

Our analysis has two parts. The empirical part employs a novel dataset from the Japan Racing Association (JRA), covering all centrally administered races from 2004 to 2023. The dataset records interim odds updates at roughly five-minute intervals from market opening to one minute before post time, allowing us to trace how odds evolve within each race. We use these data to extend the standard FLB regression framework---in which realized returns are regressed on final odds---by adding interim odds dynamics, thereby comparing horses with similar final odds but different odds trajectories.

We find a systematic association between realized returns and final-stage odds movements: horses whose odds decline in the final five minutes---that is, horses that experience a late surge in betting interest---earn significantly higher realized returns than horses with similar final odds. This pattern therefore suggests that, even among horses with similar final odds, those attracting concentrated betting interest immediately before market closure performed better ex post.


The magnitude of this path dependence is sizable relative to the conventional final-odds gradient: a 10\% final-five-minute odds movement implies a return difference substantially larger than that associated with a 10\% cross-sectional difference in final odds.\footnote{At the median final odds level of 25.5, the implied return difference associated with a 10\% final-five-minute odds movement is about 14 times larger than that associated with a 10\% cross-sectional difference in final odds.} Furthermore, we find that incorporating late-stage odds movements attenuates the negative final odds--return relationship. This attenuation suggests that the conventional FLB is partly related to the trajectory through which final odds are formed, rather than to the final odds level alone.

The second part of the analysis provides a theoretical interpretation of these empirical patterns. Building on \citet{ottaviani2009surprised,ottaviani2010noise}, we develop a two-period information-based betting model that explicitly distinguishes interim odds from final odds. In the model, final odds are formed by subsequent informed betting on top of pre-existing market shares, so a late increase in a horse's market share---equivalently, a decline in its odds---can reflect additional betting by informed bettors with favorable private signals. The model thus links the documented path dependence in returns to the process through which private information is aggregated in a parimutuel market.

The model suggests a caution for structural estimation using betting-market data. Our simulation based on the two-period information model illustrates that dynamic information aggregation can generate odds--return patterns resembling those attributed to preference distortions in static structural models of betting markets. Thus, estimates based only on final odds and realized outcomes may partly absorb informational variation into preference or belief parameters. More broadly, final odds--return patterns alone may not distinguish informational mechanisms from risk preferences, probability weighting, or subjective beliefs.

\paragraph{Related Literature} 
A wide range of mechanisms has been proposed to explain the FLB. One strand attributes the bias to preferences or probability distortions, including risk-loving behavior \citep{Weitzman}, systematic risk misperceptions \citep{jullien2000estimating, snowberg2010explaining}, and heterogeneity in both risk preferences and probability perceptions \citep{chiappori2019aggregate}. Another strand emphasizes heterogeneous beliefs under risk neutrality, showing that differences in beliefs can generate the FLB even without risk-loving preferences \citep{Ali, gandhi2015does}.

The studies most closely related to ours are \citet{ottaviani2009surprised,ottaviani2010noise}. They analyze parimutuel betting markets with imperfectly informed bettors who can condition their actions on private signals but not on final odds. Their key insight is that final odds aggregate realized private signals only after wagers have been placed, so horses attracting more signal-driven betting tend to have larger market shares, lower odds, and higher winning probabilities, generating the FLB. Our study builds on this information-based explanation and provides complementary evidence that, consistent with informed bettors' incentives to delay wagering until close to market closure, late odds declines are associated with higher realized returns among horses with similar final odds.

Several studies also emphasize that the informational efficiency of betting markets depends on institutional design. \citet{smith2006economica} show that the FLB is much weaker in Betfair's exchange environment, where bettors can post and accept wagers in a manner closer to contingent price offers than in parimutuel systems. Similarly, \citet{franck2010} provide evidence from European football betting markets that the way odds are set and adjusted affects market efficiency. These studies suggest that market efficiency depends not only on bettors' preferences or beliefs, but also on the institutional rules governing how odds are determined.

\paragraph{Structure of the Study} 
The remainder of the study is organized as follows. Section~\ref{Sec:Data} introduces the data and key empirical facts. Section~\ref{Sec:Empirical} presents the econometric framework and main empirical findings. Section~\ref{Sec:Discuss} develops the two-period betting model and discusses related implications. Section~\ref{Sec:Conclusion} concludes.

\section{Data}\label{Sec:Data}

This section describes the institutional background of Japanese horse racing, introduces our dataset, and documents key empirical patterns that motivate the subsequent econometric analysis.

\subsection{Institutional Details}

Japanese horse racing is organized by two governing bodies: the Japan Racing Association (JRA) and the National Association of Racing (NAR). The JRA administers central horse racing, known as \textit{Chuo Keiba}, at ten major racecourses, including Tokyo, Nakayama, Kyoto, and Hanshin, while the NAR oversees local races. JRA races comprise roughly 3,400 races annually and account for one of the largest horse-racing markets in the world, with recent annual betting turnover exceeding 2.8 trillion yen, or about 19 billion USD at an exchange rate of 145 yen per dollar.

This study focuses exclusively on JRA races, whose records are systematically archived in a uniform format. JRA races use a parimutuel betting system, in which odds are updated based on wager shares and final payouts are determined by the closing odds. Betting closes one minute before post time, and tickets are sold in units of 100 JPY (approximately 0.69 USD).\footnote{Races are held mainly on weekends and national holidays. Betting opens at 6:30 p.m.\ on Friday for Saturday races and at 7:30 p.m.\ on Saturday for Sunday races. Supplementary Appendix~\ref{sec:appendix_data} provides further details.}

Our analysis focuses on the win (\textit{tanshō}) pool. The JRA deducts 20\% from the total pool, so the gross payout per 1 JPY bet on horse \(i\) is
\begin{equation}
    R_{i} = \frac{(1 - 0.2)\sum_{j \in I} W_j}{W_i}
    = \frac{0.8}{s_i}, 
    \label{eqm_odds}
\end{equation}
where \(W_i\) is the amount wagered on horse \(i\), \(I\) is the set of competing horses, and \(s_i \equiv W_i / \sum_{j \in I} W_j\) is horse \(i\)'s share of total wagers.\footnote{In Japan, horse-racing odds are quoted as gross payouts inclusive of the original stake. Thus, odds of 2.5 mean that a successful 1 JPY bet pays 2.5 JPY in total, or 1.5 JPY in net profit. The JRA operates a supplemental payout scheme known as ``JRA Plus 10,'' under which a winning 100-JPY ticket with odds of 1.0 may pay 110 JPY under certain conditions. To avoid ambiguity, we exclude races in which any horse had final odds of exactly 1.0. Such cases are extremely rare, representing only \(0.0007\%\) of horses in our 20-year dataset.}  Equation~\eqref{eqm_odds} applies to both final and interim odds. Interim odds are computed from cumulative wagers placed up to each point in time, whereas payouts are determined only by the closing odds. Thus, interim odds provide snapshots of evolving market shares during the betting period, even though they do not directly determine payoffs.\footnote{One complication arises when a horse is withdrawn after betting has opened: tickets involving that horse are refunded, but posted odds may not adjust in real time. We therefore exclude all races with post-opening withdrawals to avoid discrepancies between observed odds and market expectations.}

\subsection{Horse-Race-Time-Level Panel Dataset}

Our empirical analysis relies on data from the JRA-VAN database, a licensed service that provides structured records from the JRA. For each race, the dataset contains race-level information, including racetrack, date and time, race classification, track condition, course distance, and aggregate betting volumes, as well as horse-level information such as name, age, sex, jockey, finishing position, and official time (see Supplementary Appendix~\ref{sec:appendix_data} for sample descriptive statistics). 

A distinctive feature of the JRA-VAN data is the availability of interim odds throughout the betting window. In addition to final odds published at the close of wagering, interim win odds are available at roughly five-minute intervals until one minute before post time. Because the precise timing of updates can vary slightly across races, we harmonize the data into fixed five-minute intervals. This harmonized horse-race-time-level panel allows us to trace the evolution of odds within each race and relate interim movements to final odds and race outcomes.

\subsection{Basic Facts}
We begin by documenting several key empirical patterns regarding the dynamics of odds and their relationship with realized returns.

\begin{figure}[t!]
\centering
\caption{Odds, Returns, and Wagering Activity}
\label{fg:odds_returns_wagering_activity}

\begin{subfigure}[t]{0.48\textwidth}
    \centering
    \caption{Odds versus Returns}
    \label{fg:plot_race_odds_return}
    \includegraphics[width=\textwidth]{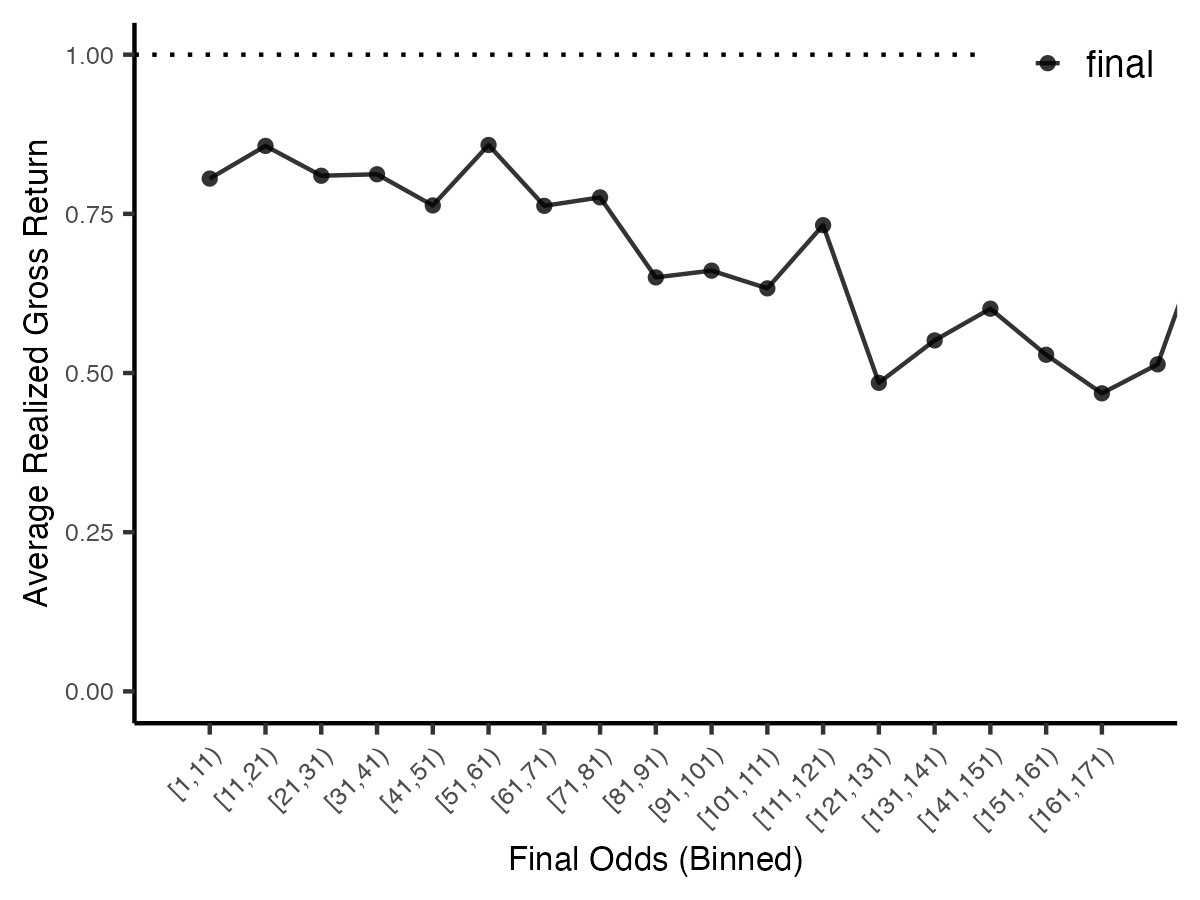}
\end{subfigure}
\hfill
\begin{subfigure}[t]{0.48\textwidth}
    \centering
    \caption{Timing of Wagering Activity}
    \label{fg:plot_race_time_share_votes}
    \includegraphics[width=\textwidth]{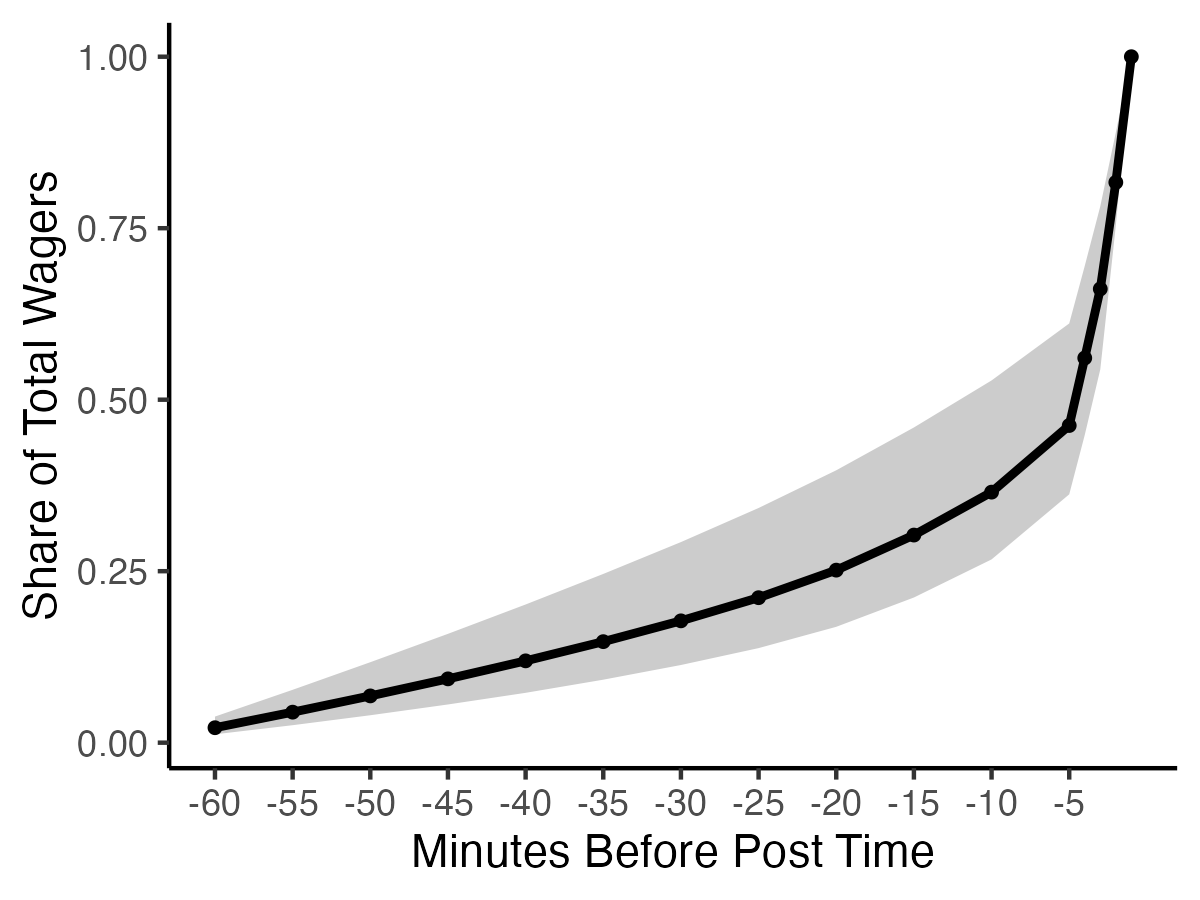}
\end{subfigure}

\begin{tablenotes}[flushleft]
\footnotesize
\item \textit{Notes:} 
Panel~(\subref{fg:plot_race_odds_return}) reports average realized gross returns by odds group ($g \in \mathcal{G}$), computed as in \eqref{eq:avreturn}. Panel~(\subref{fg:plot_race_time_share_votes}) shows the cumulative share of wagers placed from 60 minutes to post time, using five-minute intervals until five minutes before post time and one-minute intervals thereafter. Betting volume is normalized to one within each race; the solid black line and shaded region denote the median and 95\% interval across races.
\end{tablenotes}

\end{figure}

\paragraph{Realized Returns and Final Odds}
Figure~\ref{fg:plot_race_odds_return} shows average realized gross returns across fixed-width bins of final odds. Let \(\mathcal{G}\) denote the set of final-odds bins, and let \(I_g\) denote the set of horses in bin \(g\). For each bin \( g \in \mathcal{G} \), the return is computed as
\begin{equation}
\frac{1}{| I_g |} \sum_{i \in I_g} \mathbf{1}_{\{win_i = 1\}} R_i^{\ast}, \label{eq:avreturn}
\end{equation}
where \( R_i^{\ast} \) denotes the final odds, inclusive of the original stake. Losing horses yield zero returns. The figure shows that average realized returns decline with final odds, consistent with the FLB: longshots systematically underperform favorites.

\paragraph{Temporal Distribution of Bets}
Figure~\ref{fg:plot_race_time_share_votes} plots the cumulative share of total bets placed by each point before post time. The key pattern is \emph{the strong concentration of wagers just before post time}. In a typical race, only about 25\% of wagers have been placed 20 minutes before post time, while nearly half are submitted within the final five minutes.\footnote{
This concentration of late wagering parallels the ``sniping'' behavior observed in online auctions such as eBay, where participants strategically delay bids until the final moments \citep{roth2002last, ockenfels2006ebay, bajari2004ecommerce}. In the context of parimutuel betting, \citet{ottaviani2006timing} develop a theoretical framework in which informed bettors postpone their wagers to avoid revealing private information to the market. Such strategic delay results in a clustering of bets immediately before post time.
}

\paragraph{Odds Changes and Outcomes}
\begin{figure}[t!]
  \caption{Expected Return versus Interim Odds Changes}
  \label{fg:plot_expected_return_intermediate_odds_min_before_5_to_0} 
  \begin{center}
  \subfloat[-5 to 0 (final)]{\includegraphics[width = 0.4\textwidth]{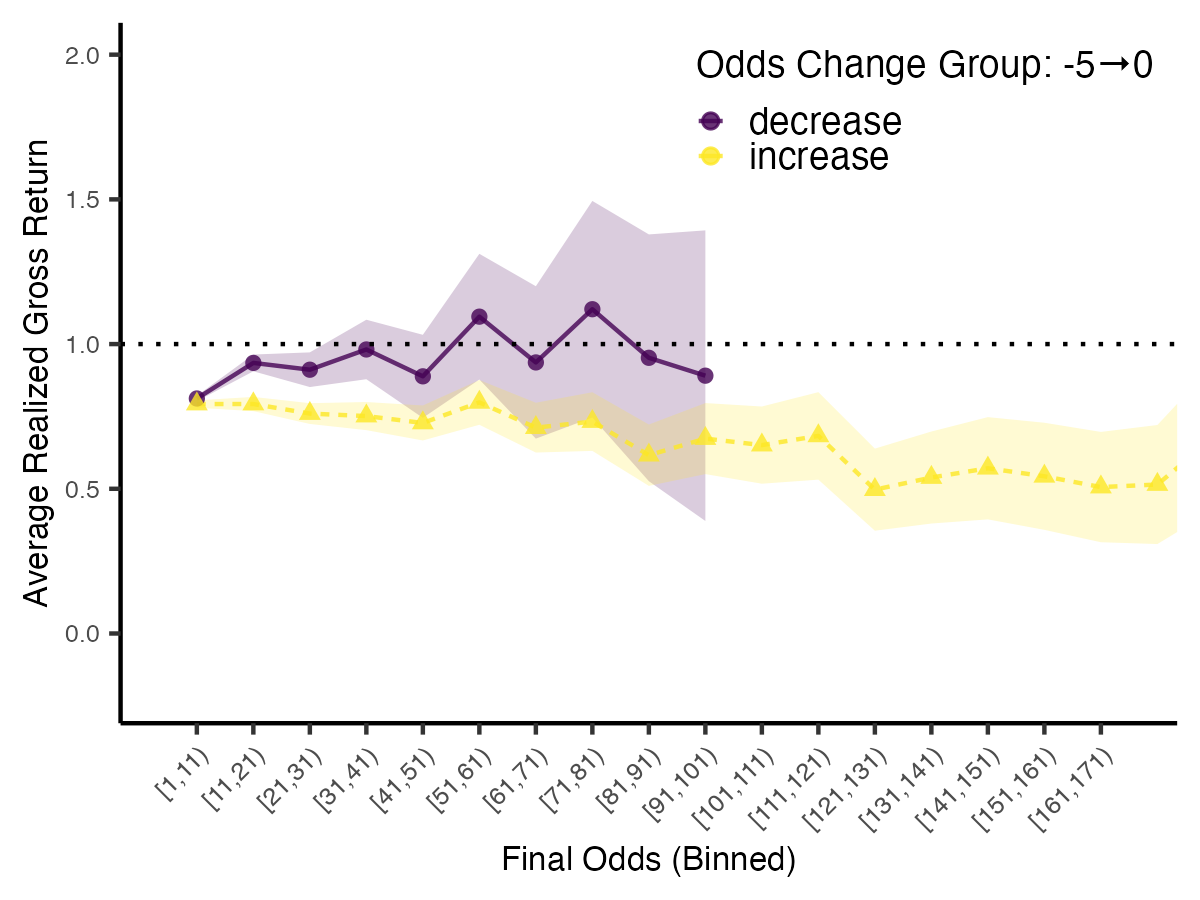}}
  \qquad\qquad
  \subfloat[-10 to -5]{\includegraphics[width = 0.4\textwidth]{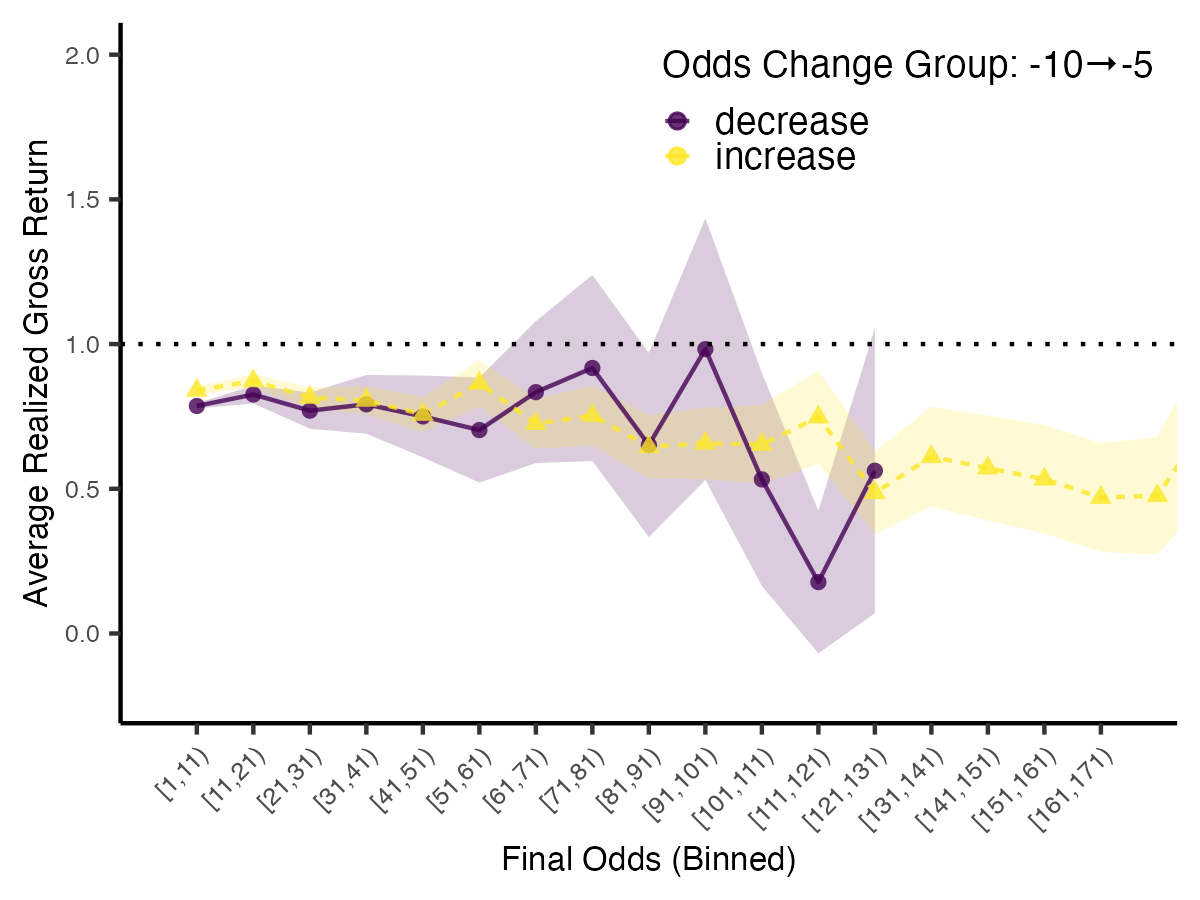}}\\
  \subfloat[-15 to -10]{\includegraphics[width = 0.4\textwidth]{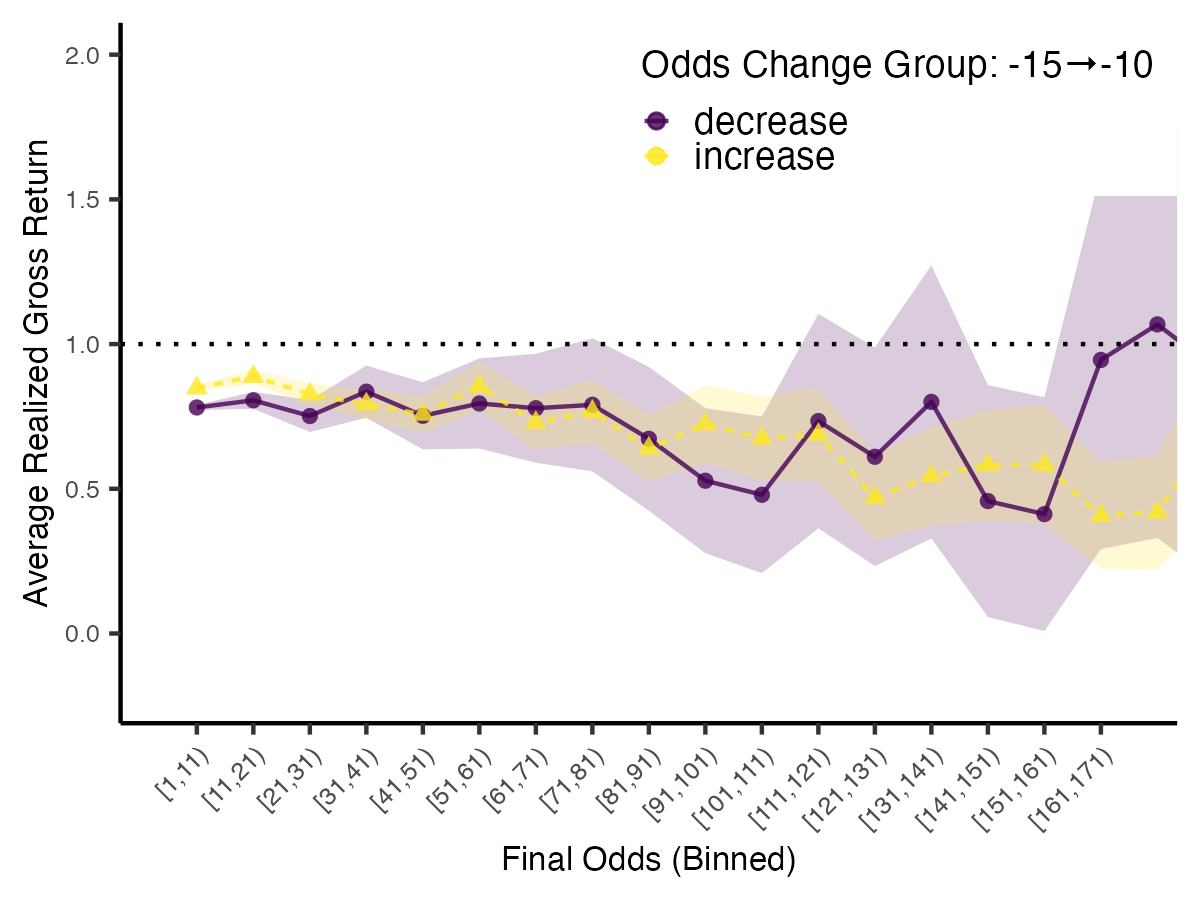}}
  \qquad\qquad
  \subfloat[-20 to -15]{\includegraphics[width = 0.4\textwidth]{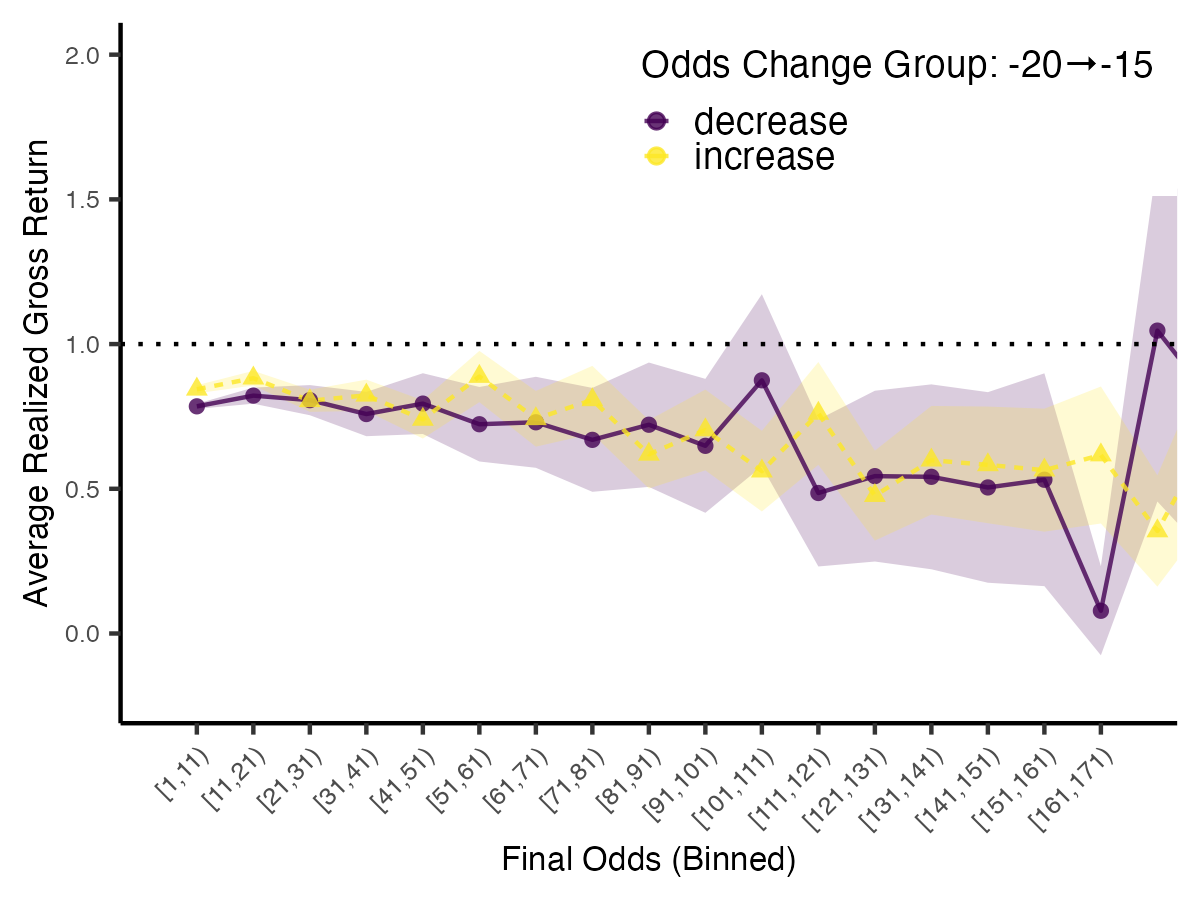}}\\
  \subfloat[-25 to -20]{\includegraphics[width = 0.4\textwidth]{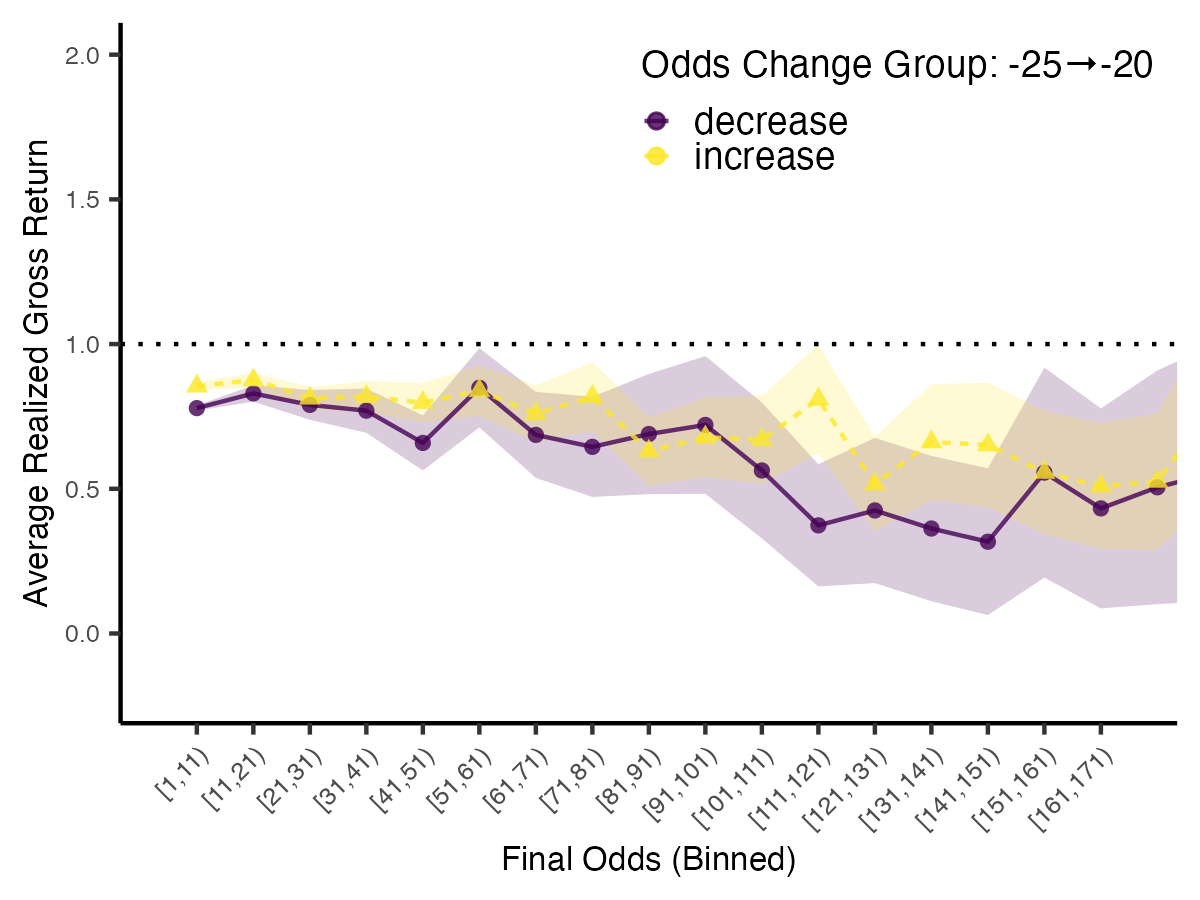}}
  \qquad\qquad
  \subfloat[-30 to -25]{\includegraphics[width = 0.4\textwidth]{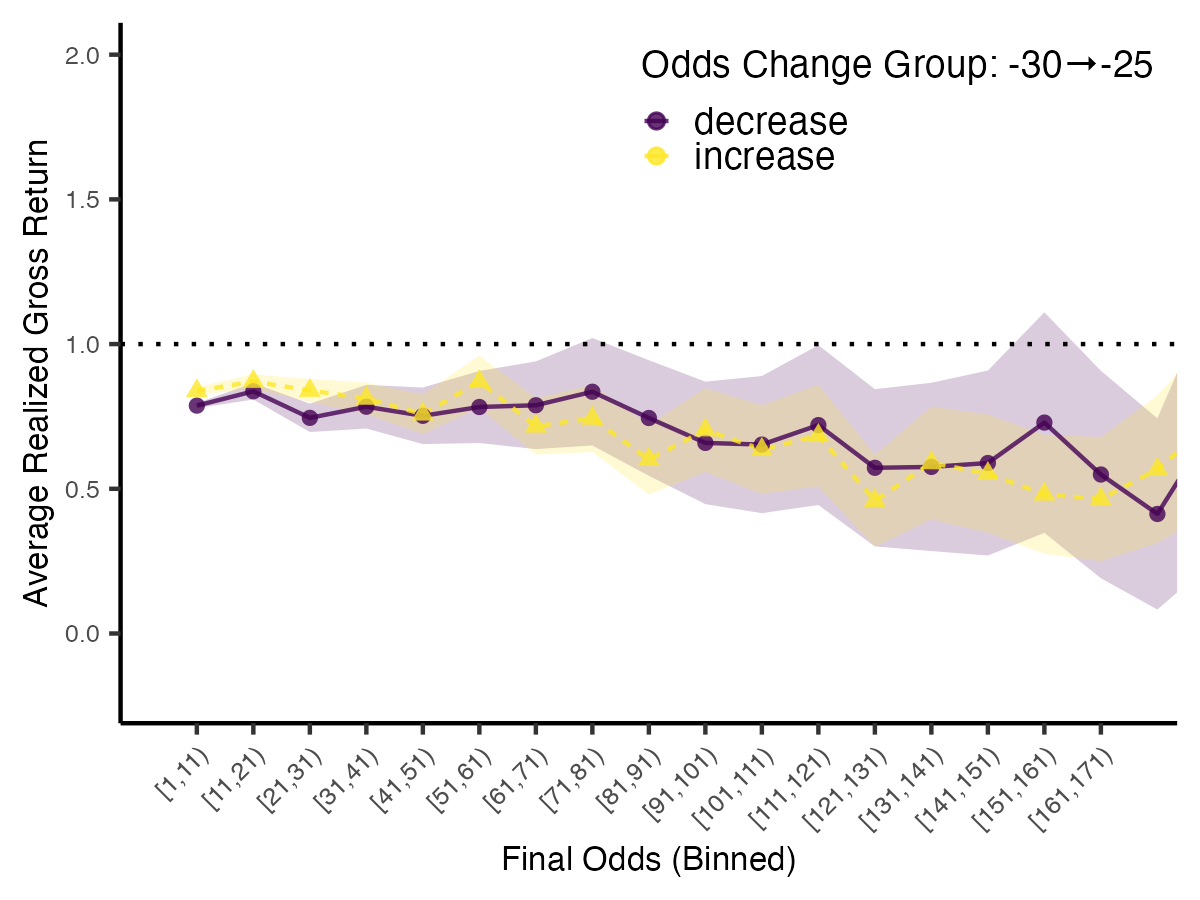}}
  
  \end{center}

\begin{tablenotes}[flushleft]
\footnotesize
\item \textit{Notes:} 
The horizontal axis represents final odds, grouped into bins of width 10. Panels (a)--(f) compare average realized returns for horses whose odds increased versus decreased during each five-minute interval before post time. Panel (a) covers the final interval, while Panels (b)--(f) examine earlier intervals: 5 to 10, 10 to 15, 15 to 20, 20 to 25, 25 to 30 minutes before post time, respectively. Shaded areas denote 95\% confidence intervals. Bins with fewer than 1,000 observations are excluded separately by group.
\end{tablenotes}

\end{figure}

We next examine whether interim odds dynamics are associated with realized returns. Figure~\ref{fg:plot_expected_return_intermediate_odds_min_before_5_to_0} compares average realized returns by final-odds bin, separately for horses whose odds increased or decreased during five-minute intervals before post time. Panel~(a) focuses on the final five-minute interval. Conditional on similar final odds, horses whose odds declined during this window---indicating a late surge in betting interest---earn consistently higher average realized returns than those whose odds increased.

Panels~(b)--(f) repeat the same comparison for earlier intervals, from 5--10 to 25--30 minutes before post time. In contrast to the final interval, odds movements in these earlier windows show no systematic association with realized returns. Thus, interim odds dynamics predict returns only near post time, highlighting the role of late-stage wagering and path dependence in betting markets.

\paragraph{Realized Portfolio Returns: Late versus Early Bettors}
To assess whether late bettors are more likely to earn positive net returns, we compare hypothetical portfolios constructed from aggregate wager shares. Let \(\lambda^{late}_{i,r}\) denote horse \(i\)'s share of final-five-minute wagers in race \(r\), and let \(s_{i,r,-5}\) denote its market share five minutes before post time. Supplementary Appendix~\ref{sec:portfolio_return} describes how \(\lambda^{late}_{i,r}\) is recovered from the data. The corresponding portfolio returns are
\[
R^{late}_r
=
\sum_i \lambda^{late}_{i,r}
\mathbf{1}_{\{win_{i,r}=1\}}R^\ast_{i,r},
\qquad
R^{early}_r
=
\sum_i s_{i,r,-5}
\mathbf{1}_{\{win_{i,r}=1\}}R^\ast_{i,r}.
\]
The late portfolio yields a gross return above one---a positive net return---in \(5.4\%\) of races, compared with \(3.9\%\) for the early portfolio.\footnote{The JRA's 20\% takeout implies a benchmark gross return of \(0.8\). The probability that the gross return lies in \((0.8,1]\) is \(46.7\%\) for the late portfolio and \(41.4\%\) for the early portfolio.}

\section{Empirical Strategy and Results}\label{Sec:Empirical}
\subsection{Econometric Strategy}\label{Sec:Strategy}
The preceding section showed that odds movements in the final five minutes before post time---when nearly half of all bets are placed---are systematically related to race outcomes. This section develops an empirical framework to quantify such path dependence in expected returns.

We begin with the baseline framework:
\begin{equation}
 \mathbf{1}_{\{win_{i}=1\}}R_{i}^{\ast}
 =
 \alpha
 + \beta R_{i}^{\ast}
 + \delta \, \text{OddsChange}_{i}
 + \gamma \, R_{i}^{\ast} \times \text{OddsChange}_{i}
 + \varepsilon_{i},
 \label{reg_main}
\end{equation}
where \( R_{i}^{\ast} \) denotes the final odds for horse \( i \), \( \text{OddsChange}_{i} \) is a variable or vector of variables summarizing the trajectory of interim odds, and \( \varepsilon_i \) is a mean-zero idiosyncratic error term. The dependent variable is the realized gross return to a 1 JPY bet, equal to \( R_i^\ast \) if horse \( i \) wins and zero otherwise. Hence,
\[
\mathbb{E}\!\left[
\mathbf{1}_{\{win_{i}=1\}}R_{i}^{\ast}
\mid R_{i}^{\ast}, \text{OddsChange}_{i}
\right]
=
\alpha
+ \beta R_{i}^{\ast}
+ \delta \, \text{OddsChange}_{i}
+ \gamma \, R_{i}^{\ast} \times \text{OddsChange}_{i},
\]
where \( \mathbb{E}[\cdot \mid X] \) denotes the expectation conditional on covariates \( X \).

This specification highlights three parameters of interest. The coefficient \(\beta\) captures how expected returns vary with final odds, thereby testing for the conventional FLB. The coefficient \(\delta\) measures whether the trajectory of interim odds predicts expected returns conditional on final odds. Finally, \(\gamma\) allows this relationship to vary with the level of final odds, capturing heterogeneity in path dependence across the odds distribution.

Under full-information rational expectations and risk neutrality, expected gross returns should be uniformly \(0.8\), reflecting the 20\% JRA takeout rate. This null hypothesis implies \(\alpha=0.8\) and \(\beta=\delta=\gamma=0\).

\subsection{Main Results}\label{Sec:Results}

\begin{table}[h!]
  \footnotesize
  \caption{Estimation Results with Additional Fixed Effects and Controls}
  \label{tab:main_results}
  \begin{center}
      
\begin{tabular}[t]{lcccccc}
\toprule
  & (1) & (2) & (3) & (4) & (5) & (6)\\
\midrule
$R_{i}^{*}$ & -0.0014 & -0.0010 & -0.0010 & -0.0012 & -0.0012 & -0.0009\\
 & (0.0001) & (0.0002) & (0.0002) & (0.0002) & (0.0002) & (0.0002)\\
$\frac{\Delta R_{i,[-5, 0]}}{R_{i,-5}}$ &  & -0.3386 &  &  &  & -0.4201\\
 &  & (0.0392) &  &  &  & (0.0383)\\
$\frac{\Delta R_{i,[-10, 0]}}{R_{i,-10}}$ &  &  & -0.1559 &  &  & \\
 &  &  & (0.0286) &  &  & \\
$\frac{\Delta R_{i,[-15, 0]}}{R_{i,-15}}$ &  &  &  & -0.0946 &  & \\
 &  &  &  & (0.0248) &  & \\
$\frac{\Delta R_{i,[-20, 0]}}{R_{i,-20}}$ &  &  &  &  & -0.0615 & \\
 &  &  &  &  & (0.0207) & \\
$\frac{\Delta R_{i,[-10, -5]}}{R_{i,-10}}$ &  &  &  &  &  & 0.1612\\
 &  &  &  &  &  & (0.0621)\\
$\frac{\Delta R_{i,[-15, -10]}}{R_{i,-15}}$ &  &  &  &  &  & 0.3674\\
 &  &  &  &  &  & (0.0717)\\
$\frac{\Delta R_{i,[-20, -15]}}{R_{i,-20}}$ &  &  &  &  &  & 0.2183\\
 &  &  &  &  &  & (0.0631)\\
$R_{i}^{*} \times \frac{\Delta R_{i,[-5, 0]}}{R_{i,-5}}$ &  & 0.0000 &  &  &  & 0.0005\\
 &  & (0.0003) &  &  &  & (0.0003)\\
$R_{i}^{*} \times \frac{\Delta R_{i,[-10, 0]}}{R_{i,-10}}$ &  &  & -0.0001 &  &  & \\
 &  &  & (0.0003) &  &  & \\
$R_{i}^{*} \times \frac{\Delta R_{i,[-15, 0]}}{R_{i,-15}}$ &  &  &  & 0.0000 &  & \\
 &  &  &  & (0.0002) &  & \\
$R_{i}^{*} \times \frac{\Delta R_{i,[-20, 0]}}{R_{i,-20}}$ &  &  &  &  & 0.0000 & \\
 &  &  &  &  & (0.0002) & \\
$R_{i}^{*} \times \frac{\Delta R_{i,[-10, -5]}}{R_{i,-10}}$ &  &  &  &  &  & -0.0009\\
 &  &  &  &  &  & (0.0007)\\
$R_{i}^{*} \times \frac{\Delta R_{i,[-15, -10]}}{R_{i,-15}}$ &  &  &  &  &  & -0.0007\\
 &  &  &  &  &  & \vphantom{1} (0.0005)\\
$R_{i}^{*} \times \frac{\Delta R_{i,[-20, -15]}}{R_{i,-20}}$ &  &  &  &  &  & -0.0008\\
 &  &  &  &  &  & (0.0005)\\
\midrule
Num.Obs. & 894127 & 894127 & 894127 & 894127 & 894127 & 894127\\
R2 & 0.002 & 0.002 & 0.002 & 0.002 & 0.002 & 0.002\\
R2 Adj. & 0.001 & 0.001 & 0.001 & 0.001 & 0.001 & 0.001\\
FE: Race-Year & X & X & X & X & X & X\\
FE: Race-Location & X & X & X & X & X & X\\
FE: Race-Grade & X & X & X & X & X & X\\
FE: Jockey & X & X & X & X & X & X\\
Horse Characteristics & X & X & X & X & X & X\\
\bottomrule
\end{tabular}

  \end{center}
     \begin{tablenotes}[flushleft]
\footnotesize
\item \textit{Notes:} All regressions include fixed effects for year, racetrack location, race grade, horse, and jockey, as well as controls for horse age, horse sex, race type, and race distance. Standard errors are clustered at the race level.
    \end{tablenotes}
\end{table}

Table~\ref{tab:main_results} reports the main estimation results. Our main specification is Column~(2), which focuses on odds movements during the final five minutes before post time. To measure these movements, we define the odds-change rate over a window from \(\tau\) minutes before post time to the final odds as
\[
\frac{\Delta R_{i,[-\tau,0]}}{R_{i,-\tau}}
\equiv
\frac{R_i^\ast - R_{i,-\tau}}{R_{i,-\tau}},
\]
where \(R_i^\ast\) denotes the final odds and \(R_{i,-\tau}\) denotes the interim odds for horse \(i\) observed \(\tau>0\) minutes before post time. A negative value of this variable indicates that the horse's odds declined over the window, corresponding to an increase in betting interest.

For the main five-minute specification, we estimate
\begin{equation}
 \mathbf{1}_{\{win_i=1\}}R_i^\ast
 =
 \alpha
 + \beta R_i^\ast
 + \delta
 \frac{\Delta R_{i,[-5,0]}}{R_{i,-5}}
 + \gamma
 R_i^\ast
 \times
 \frac{\Delta R_{i,[-5,0]}}{R_{i,-5}}
 + Z_i'\zeta
 + \varepsilon_i ,
 \label{reg_main2}
\end{equation}
where \(Z_i\) denotes a vector of additional covariates and fixed effects, including race-year, racetrack-location, race-grade, and jockey fixed effects, as well as horse-level characteristics such as age and sex, race type, and race distance. These controls absorb a wide range of observed characteristics and fixed heterogeneity across races, horses, and jockeys.

Column~(1) reports a restricted specification without path dependence, imposing \( \delta=\gamma=0 \). The coefficient on final odds is negative and statistically significant, with \( \hat{\beta}=-0.0014 \), implying a 0.14 JPY decline in expected returns for a standard 100 JPY wager when final odds increase by one unit. This rejects the risk-neutral rational-expectations benchmark of constant expected returns and reproduces the conventional FLB.

Column~(2) adds the final-five-minute odds-change rate, which is the main variable of interest. The coefficient on this variable is negative and statistically significant, with \(\hat{\delta}=-0.3386\). This estimate indicates that, among horses with similar final odds, those whose odds fall in the final five minutes tend to yield higher realized returns. This pattern is consistent with late wagering containing information about subsequent performance. The interaction term with \(R_i^\ast\) is close to zero, with \(\hat{\gamma}=0.0000\), suggesting that this relationship does not vary substantially across the final-odds distribution.

Notably, once late-stage odds movements are included, the coefficient on final odds decreases in magnitude from \(\hat{\beta}=-0.0014\) in Column~(1) to \(\hat{\beta}=-0.0010\) in Column~(2). This attenuation suggests that part of the negative association between final odds and realized returns is related to late-stage odds dynamics, rather than to the final odds level alone.

Columns~(3)--(5) extend the same odds-change measure to longer windows ending at post time: 10, 15, and 20 minutes, respectively. The estimated coefficients remain negative but become smaller in magnitude as the window length increases. This pattern suggests that the association between odds movements and expected returns is strongest near post time. Column~(6), which includes adjacent five-minute odds changes simultaneously, confirms that the negative association is concentrated near market closure: the final-five-minute coefficient remains negative, whereas the earlier adjacent intervals have positive coefficients.\footnote{Supplementary Appendix~\ref{appen_path_dependence} further examines adjacent five-minute paths. The results show that final-five-minute odds declines are associated with particularly high returns when they reverse a preceding odds increase, while final-five-minute odds increases are associated with particularly low returns when they reverse a preceding odds decline.}


In summary, the regression results indicate that expected returns are related not only to the level of final odds, as emphasized by the FLB, but also to the trajectory through which those odds are reached. In particular, odds declines in the final minutes before post time are associated with higher realized returns among horses with similar final odds. This finding suggests that late-stage odds movements contain information that is not fully summarized by the final odds level alone.

Several points are worth noting. First, the main results are robust to alternative control specifications, including specifications without controls and with race fixed effects, as shown in Supplementary Appendix~\ref{sec:robustness_fe}. Second,  as a supplementary analysis, Supplementary Appendix~\ref{appen_last_five_minute_decomposition} decomposes the final-five-minute odds change into five-to-two-minute and two-minute-to-final changes for the subsample of races with an interim odds update at the two-minute mark.\footnote{Because this decomposition requires an interim odds update at the two-minute mark, it can be implemented only for the subsample of races with such updates; we therefore treat it as a supplementary analysis rather than as the main specification.} Both components are negatively associated with realized returns, but the association is stronger in the final segment, suggesting that the most recent odds revisions near market closure play a particularly important role. Third, these associations should not be interpreted as evidence of \emph{ex-ante} return predictability. They are \emph{ex-post} associations, because bettors cannot condition their wagers on final odds once they are realized. We discuss this issue further in Supplementary Appendix~\ref{App:ExAntePredict}. The remainder of this subsection discusses two additional issues that help interpret the economic and informational content of late-stage odds movements.

\paragraph{Magnitude of Last-Minute Odds Movements}
To gauge the economic magnitude of this path dependence, we compare two effects evaluated at the same final-odds level \(R^{\ast}\). The first is the expected-return difference associated with a 10\% cross-sectional difference in final odds:
\[
\Delta_{\text{final}}(R^{\ast})
=
\hat{\beta} \times 0.1 R^{\ast}.
\]
The second is the expected-return difference associated with a 10\% odds movement during the final five minutes:
\[
\Delta_{\text{path}}(R^{\ast})
=
\hat{\delta} \times 0.1
+
\hat{\gamma} R^{\ast} \times 0.1.
\]
Table~\ref{tab:info_magnitude} reports these magnitudes at the 25th, 50th, and 75th percentiles of final odds. Across these points, \(\Delta_{\text{path}}\) is consistently larger in absolute value than \(\Delta_{\text{final}}\). At the median final-odds level, for example, the implied return difference associated with a 10\% final-five-minute odds movement is about 14 times larger than that associated with a 10\% cross-sectional difference in final odds.

\begin{table}[t!]
  \footnotesize
  \caption{Magnitude: Final Odds versus Last-Five-Minute Odds Changes}
  \label{tab:info_magnitude}
  \begin{center}
      
\begin{tabular}[t]{lccc}
\toprule
Final odds $R^{\ast}$ & 8.6 & 25.5 & 82.7\\
\midrule
$\Delta_{\text{final}}(R^{\ast})$ & -0.0008 & -0.0024 & -0.0079\\
$\Delta_{\text{path}}(R^{\ast})$ & -0.0338 & -0.0338 & -0.0336\\
$\Delta_{\text{path}}(R^{\ast})/\Delta_{\text{final}}(R^{\ast})$ & 41.05 & 13.82 & 4.24\\
\bottomrule
\end{tabular}

  \end{center}
\begin{tablenotes}[flushleft]
\footnotesize
\item \textit{Notes:} The final-odds levels \(R^{\ast}\) correspond to the 25th, 50th, and 75th percentiles of the final-odds distribution. The table compares the expected-return implications of a 10\% cross-sectional difference in final odds, \(\Delta_{\text{final}}(R^{\ast})=\hat{\beta}\times 0.1R^{\ast}\), and a 10\% final-five-minute odds movement, \(\Delta_{\text{path}}(R^{\ast})=\hat{\delta}\times 0.1+\hat{\gamma}R^{\ast}\times 0.1\). Estimates are based on the benchmark specification with the same controls and fixed effects as in Table~\ref{tab:main_results}.
\end{tablenotes}
\end{table}

\paragraph{Asymmetry between Odds Increases and Declines}
We also examine whether the association between last-five-minute odds changes and realized returns is symmetric between odds increases and declines. Supplementary Appendix~\ref{sec:robustness_asymmetry} decomposes the odds-change variable into positive and negative components. The estimates point in the expected directions and indicate a stronger association for odds declines, especially at higher final odds. This asymmetry is consistent with Panel~(a) of Figure~\ref{fg:plot_expected_return_intermediate_odds_min_before_5_to_0}, where horses with final-five-minute odds declines tend to exhibit higher realized returns, particularly in the middle range of final odds.

\section{Theoretical Interpretation and Additional Implications}\label{Sec:Discuss}

The preceding empirical analysis documents path dependence in expected returns: among horses with similar final odds, those whose odds decline shortly before post time tend to yield higher realized returns. This section provides a simple information-based interpretation of this pattern in a parimutuel betting market.

We develop a two-period extension of the information-based betting model of \citet{ottaviani2009surprised,ottaviani2010noise}. In their simultaneous-betting framework, informed bettors place bets based on private signals about the winning horse, and final odds are determined by the resulting aggregate betting shares. We modify this benchmark by allowing some bets to be placed before informed bettors arrive. Final market shares, and hence final odds, are then determined by the combination of these pre-existing bets and subsequent informed betting.

We consider a race with \(K\in\mathbb{N}_{+}\) horses and objective winning probabilities \(q=(q_1,\ldots,q_K)\). Before informed bettors arrive, \(B_0 \geq 0\) first-period bets have already been placed, generating market shares \(s=(s_1,\ldots,s_K)\), so horse \(m\) has \(B_0s_m\) pre-existing bets. In the second period, \(M>0\) informed bettors observe the first-period market shares, equivalently the interim odds, and receive private posterior beliefs \(p\). They then choose which horse to bet on, knowing that other informed bettors draw beliefs from the same distribution. 


The expected net return from a unit bet on a horse with final market share \(\pi\) is
\[
\overline R_{\pi}
=
\frac{1-\rho}{\pi}\Pr(\text{winner}\mid\pi)-1,
\]
where \(\rho\) is the takeout rate. In the two-period model without abstention, if \(x\) second-period bettors choose horse \(m\), its final market share is
\(\pi=(B_0s_m+x)/(B_0+M)\), where we assume that \(B_0s_m\) is an integer. Equivalently, for a given final market share \(\pi\), the implied number of second-period bettors choosing horse \(m\) is $
x_m(\pi)=(B_0+M)\pi-B_0s_m$. Following the logic of Proposition~8 of \citet{ottaviani2010noise}, Bayes' rule yields
\[
\overline R_{\pi}
=
\frac{1-\rho}{\pi}
\frac{
\sum_{m\in\mathcal M(\pi)}
q_m
\binom{M}{x_m(\pi)}
\eta(m\mid m)^{x_m(\pi)}
\left[1-\eta(m\mid m)\right]^{M-x_m(\pi)}
}{
\sum_{m\in\mathcal M(\pi)}
\sum_{\ell=1}^{K}
q_\ell
\binom{M}{x_m(\pi)}
\eta(m\mid \ell)^{x_m(\pi)}
\left[1-\eta(m\mid \ell)\right]^{M-x_m(\pi)}
}
-1,
\]
where \(\mathcal M(\pi)=\{m:x_m(\pi)\in\{0,1,\ldots,M\}\}\) is the set of horses for which \(\pi\) is feasible, and \(\eta(m\mid \ell)\) denotes the equilibrium probability that an informed bettor chooses horse \(m\) conditional on horse \(\ell\) being the true winner.


\begin{figure}[t!]
\caption{Model-Implied Expected Returns by Final Market Share and Share Movement}
  \label{fg:plot_theoretical_market_share_expected_return} 
  \begin{center}
  \subfloat[One- vs. Two-Period Expected Returns
\label{fg:one_vs_two_period_return_curve}]{
    \includegraphics[width = 0.48\textwidth]{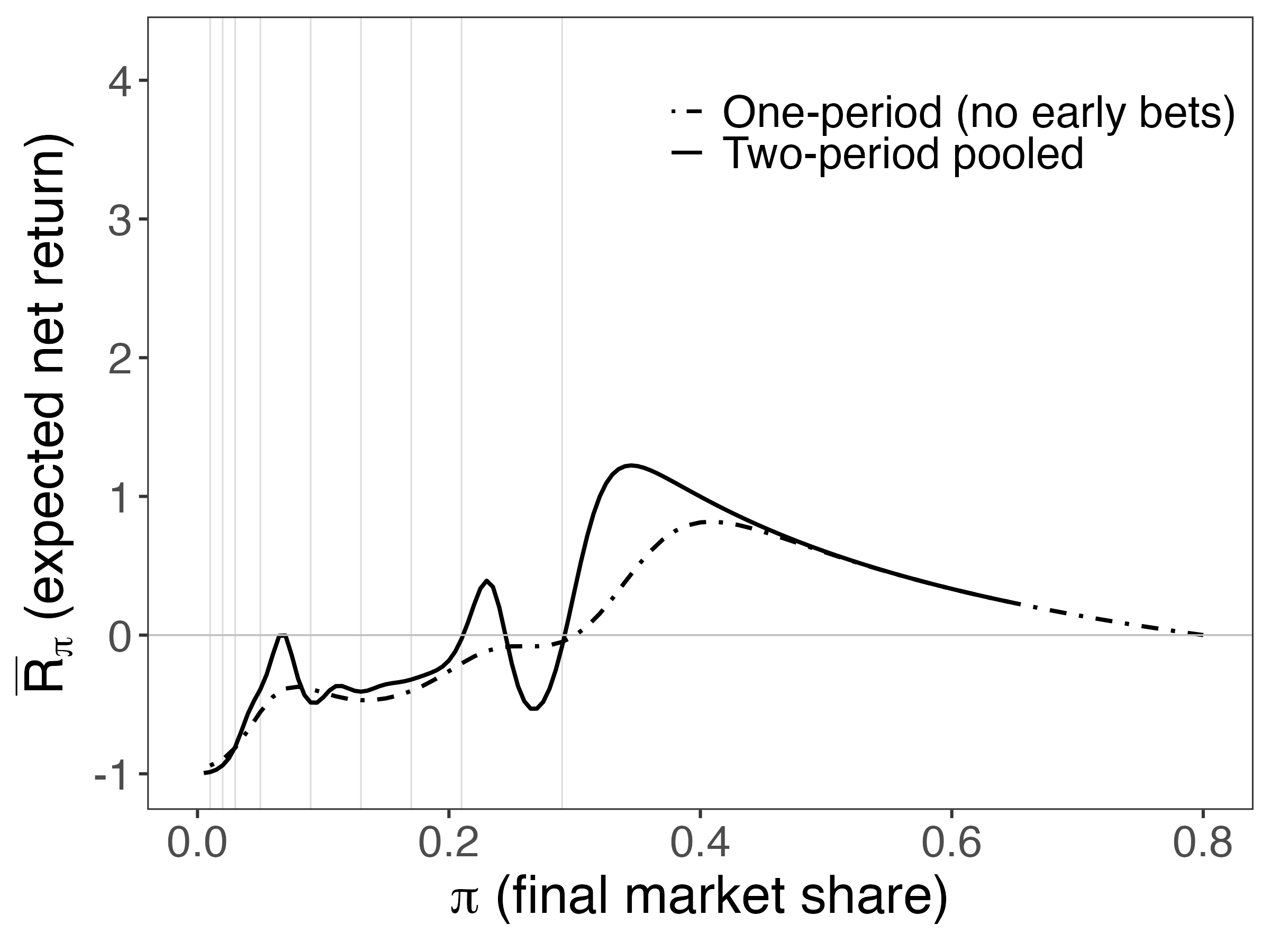}
  }
  \subfloat[Expected Returns by Share Movement\label{fg:share_change_second_period}]{
    \includegraphics[width = 0.48\textwidth]{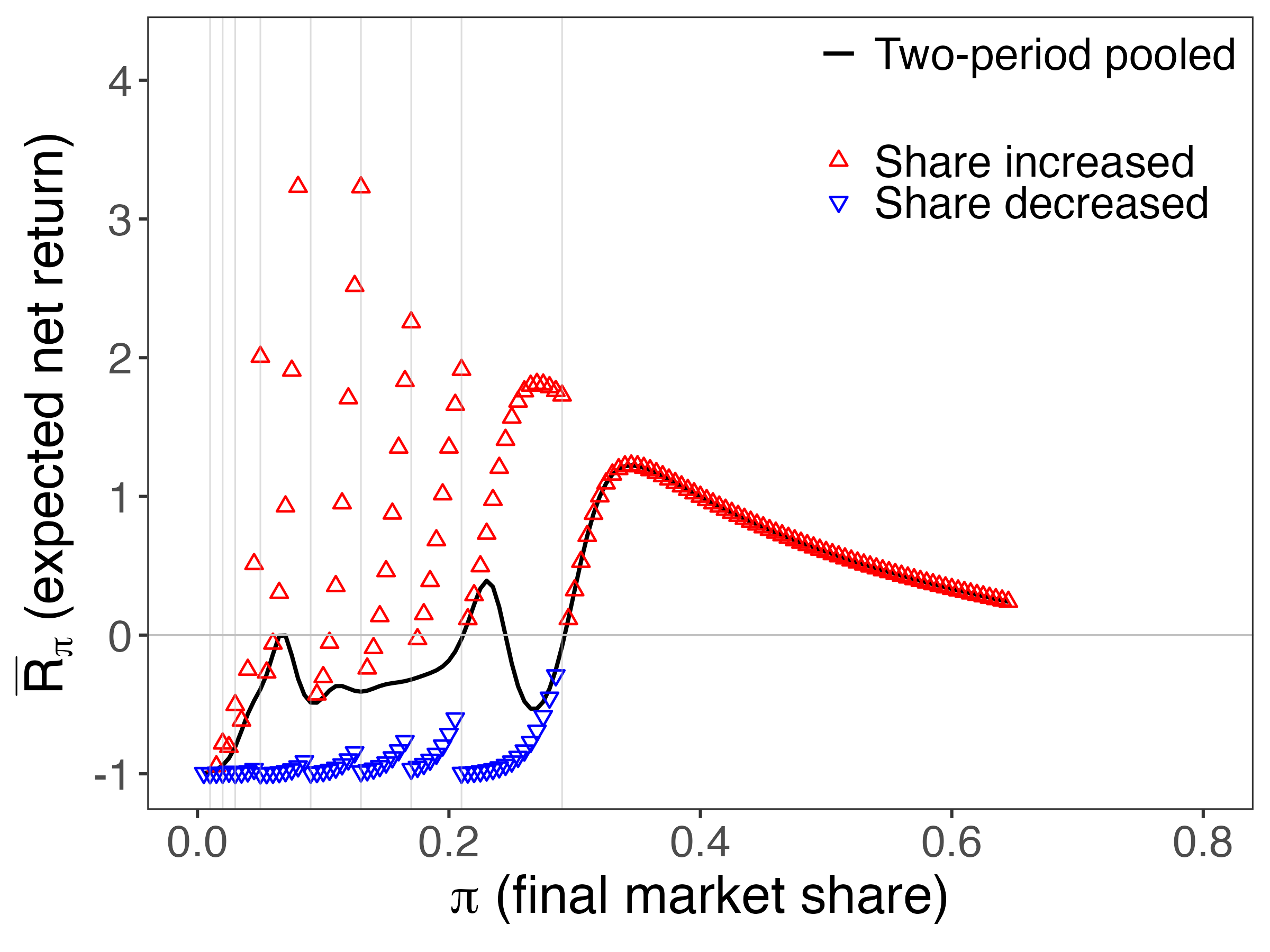}
  }  
  \end{center}
  \footnotesize
  \textit{Notes:} The figure plots the expected (net) return $\overline{R}_{\pi}$ against the final market share $\pi$ for a nine-horse race ($K=9$) with objective winning probabilities 
  $q=(0.29,0.21,0.17,0.13,0.09,0.05,0.03,0.02,0.01)$ and a takeout rate of 20\%. Panel (a) compares a one-period benchmark with 100 informed bettors who bet simultaneously to a two-period model with 100 pre-existing first-period bets and 100 informed second-period bettors ($M=100$). The nine vertical gray lines mark the first-period market shares, which are equal to $q$. Conditional on winner \(k\), signals are drawn from \(\mathrm{Dirichlet}(\theta q+e_k)\), with \(\theta=100\). In Panel~(b), each triangle represents a model-implied conditional expected return for a given final market share \(\pi\) and a given direction of movement from the first-period share. Upward triangles correspond to cases in which the final market share exceeds the first-period share, i.e., odds decline; downward triangles correspond to cases in which the final market share is below the first-period share, i.e., odds increase.
\end{figure}


Figure~\ref{fg:plot_theoretical_market_share_expected_return} reports the resulting expected-return curves for a nine-horse race (i.e., $K=9$) with objective winning probabilities \(q=(0.29,0.21,0.17,0.13,0.09,0.05,0.03,0.02,0.01)\), following the numerical example and Dirichlet signal structure in \citet{ottaviani2010noise}.\footnote{Specifically, conditional on horse \(k\) being the winner, private posterior beliefs are drawn from \(p\mid k \sim \mathrm{Dirichlet}(\theta q+e_k)\), where \(e_k\) is the \(k\)-th unit vector and \(\theta=100\).} Panel~(a) compares the one-period benchmark, in which all 100 informed bettors bet simultaneously (i.e., $M=100$), with the two-period model, in which 100 first-period bets are already in place before 100 informed second-period bettors arrive. In the two-period model, the first-period market shares are set equal to the objective probabilities, \(s=q\).\footnote{We examine cases in which first-period market shares are biased in Supplementary Appendix~\ref{App:BiasedFirstPeriodShares}.} The one-period benchmark generates expected returns that are monotonically increasing in final market shares, or equivalently decreasing in final odds, over the relevant range \((\pi<0.5)\). By contrast, the two-period model produces a broadly increasing but non-monotonic expected-return curve with sizable local fluctuations.

These fluctuations follow a systematic pattern: expected returns are higher when a horse's final market share rises above its first-period share---that is, to the right of the corresponding vertical line---meaning that its share increases and its odds decline during the second period, and lower when the final market share falls below its first-period share---that is, to the left of the corresponding vertical line. Panel~(b) makes this pattern explicit by plotting expected returns conditional not only on the final market share, but also on whether the market share increases or decreases from the first period to the final market. For the same final market share, expected returns are higher when the share increases during the second period and lower when it declines.

This implication is consistent with the empirical evidence documented in the previous section. In the dynamic setting, private information held by informed bettors is incorporated into market shares through betting activity close to post time. As a result, even conditional on final odds, the path by which those odds are reached can contain information about subsequent realized returns. The model therefore suggests that an information-based mechanism may account not only for the static FLB in final odds, but also for the path-dependent association between late odds movements and realized returns.

\paragraph{Additional Testable Implications}
In addition to rationalizing the main empirical finding, the model points to several additional implications, some of which echo those discussed in \citet{ottaviani2010noise}. We focus on three dimensions: signal precision, late betting participation, and the number of possible outcomes.

The first implication concerns the precision parameter \(\theta\) of the Dirichlet signals. Lower signal precision weakens the informativeness of private signals and is therefore expected to attenuate the FLB. We examine this implication in Supplementary Appendix~\ref{sec:add_test_racetrack} by comparing races across racetrack conditions, which proxy for the precision of private information. The results show that the final-odds gradient is attenuated under adverse track conditions: the FLB is substantially weakened, and its point estimate even reverses under the worst track conditions. This pattern is consistent with the prediction that noisier private information reduces the strength of the bias.\footnote{The corresponding evidence for late-stage odds movements is directionally similar, although statistically weaker.}

The second implication concerns the participation margin of informed bettors. In the two-period model, the information incorporated into final odds may depend on the extent of late betting activity. In Supplementary Appendix~\ref{sec:add_test_participant}, we use the share of bets placed in the final five minutes as a proxy for this margin and compare races with high and low final-five-minute betting shares. The results show that both the FLB and the association between odds declines and realized returns are weaker when the final-five-minute betting share is high. Although this pattern may appear inconsistent with the idea that more late betting strengthens information aggregation, it can be rationalized by selection into participation. As discussed by \citet{ottaviani2010noise}, weak-signal bettors may abstain when the private value of betting is finite. Low late participation may therefore imply stronger selection on signal strength, whereas high late participation may include more weak-signal bettors and dilute the information content of late odds movements. Because our data observe realized betting activity but not the pool of potential informed bettors, we interpret this exercise as suggestive rather than as a direct test of the participation channel.

The third implication concerns the number of possible outcomes. When the outcome space is large relative to betting volume, final odds become less informative and the FLB should weaken. Supplementary Appendix~\ref{App:Quinella} examines this prediction in the quinella (\textit{umaren}) market, where bettors select an unordered pair of horses.\footnote{A quinella ticket pays if the selected pair finishes first and second in either order. Unlike exacta bets, quinella bets do not require the correct order of finish.} The results show that the negative odds--return relationship remains but is substantially weaker than in the win market.

\paragraph{Implications for Risk-Preference Estimation Using Betting-Market Data}

A related implication concerns the use of betting-market or prediction-market data to infer risk preferences. The FLB is often interpreted as evidence of risk-loving behavior or probability weighting, and recent structural studies estimate preferences and beliefs from final odds and realized outcomes \citep[e.g.,][]{gandhi2015does,chiappori2019aggregate}. The information-based mechanism studied in this paper raises a potential misspecification concern: if final odds reflect not only preferences and beliefs but also private-information aggregation and the timing of informed betting, estimates that omit this channel may partly attribute informational variation to preferences or subjective probability distortions.

To illustrate this concern, we conduct a simulation exercise based on the two-period betting model. We consider nine-horse races, randomize objective winning probabilities across races, and generate final odds and realized outcomes from the model. We then apply the estimation approach of \citet{chiappori2019aggregate} to the simulated data. This exercise asks whether data generated solely by an information-based mechanism could give rise to estimates that would be interpreted as preference or belief distortions in a structural model that does not explicitly account for information aggregation (see Supplementary Appendix~\ref{App:RiskPreferenceSimulation} for the simulation design and estimation details).

The results show that the homogeneous rank-dependent expected utility specification provides the best fit to the simulated data. Since the simulated data are generated without imposing rank-dependent preferences, this finding suggests that dynamic information aggregation can generate odds--return patterns that are observationally similar to those attributed to preference distortions in final-odds-based structural models. This is noteworthy because \citet{chiappori2019aggregate} also find that the homogeneous rank-dependent expected utility model best fits their field data. Our simulation therefore suggests a possible reinterpretation of this finding: part of the empirical success of rank-dependent expected utility may reflect an omitted information channel, rather than risk preferences and probability weighting alone.

\section{Conclusion}\label{Sec:Conclusion}

This paper studies information aggregation in parimutuel betting markets using interim odds observed throughout the betting period. We show that odds trajectories contain information beyond final odds alone: among horses with similar final odds, those whose odds decline in the final minutes before post time tend to yield higher realized returns. 

We interpret this path dependence through a simple two-period information-based betting model. In the model, final odds are formed by subsequent informed betting on top of pre-existing market shares. A late increase in a horse's market share---equivalently, a decline in its odds---therefore reflects additional betting induced by informed bettors' private signals. The model links path dependence in odds to information aggregation.

Beyond horse-race betting, our findings point to a broader empirical issue: market prices may confound preferences, beliefs, and information aggregation. Our simulation exercise suggests that information aggregation alone can generate odds--return patterns resembling those attributed to rank-dependent preferences. Final odds--return patterns alone may therefore be insufficient to distinguish these mechanisms, while price dynamics such as interim odds may provide useful variation for separating them. More broadly, the results raise a theoretical question about how market structure shapes information aggregation. Parimutuel markets, where final odds are determined only after the betting window closes and contingent orders are unavailable, may differ from asset markets, where prices are formed through market-clearing mechanisms. Understanding how trading institutions shape the information content of prices is therefore an important direction for future research.

\newpage
\setstretch{1.0}
\bibliographystyle{aer}
\bibliography{horseracing}

\newpage
\appendix

\setstretch{1.1}
\begin{center}
  \huge{\textbf{Supplementary Appendix}} 
\end{center}

\section{Data Details}\label{sec:appendix_data}
 \renewcommand\theequation{\thesection.\arabic{equation}}
 \renewcommand\thefigure{\thesection.\arabic{figure}} 
  \renewcommand\thetable{\thesection.\arabic{table}} 
  \setcounter{equation}{0}
 \setcounter{figure}{0}
  \setcounter{table}{0}

\paragraph{Betting Channels in JRA Races}
Wagers on JRA races can be placed through a variety of channels, including online platforms and physical betting venues. The primary online channel is Internet Programmed Automatic Transmission (IPAT), which requires prior registration and linkage to a domestic bank account. Another online option, JRA Direct, operates via credit card and allows same-day registration and betting, though it is accessible only via desktop browsers and does not support mobile devices. In addition to these online channels, bets may also be placed in person at JRA-operated racecourses and WINS (off-track betting facilities).

Although the JRA does not publish micro-level data on individual bettors, industry reports and publicly available aggregate statistics indicate that online channels---namely IPAT and JRA Direct---collectively account for over 70\% of total wagering turnover. Betting at physical locations tends to be dominated by older bettors and is often characterized by larger per-bet amounts. In contrast, online platforms attract a broader demographic and are typically associated with smaller but more frequent wagers.

The cutoff times for purchasing JRA betting tickets vary depending on the purchase method:
\begin{enumerate}
    \item JRA IPAT: Up to 1 minute before the race starts.
    \item At the racecourse or WINS (off-track betting facilities): Up to 2 minutes before the race starts.
    \item JRA Direct and telephone betting: Up to 5 minutes before the race starts.
\end{enumerate}

\paragraph{Summary Statistics}
Table~\ref{tb:summary_statistics_horse_race} reports summary statistics for win-odds data from 2004 to 2023 at both the horse-race level and the horse-race-time level. Panels~(a) and (b) present basic characteristics of races and individual horses, respectively. On average, horses are 3.6 years old, each race features an average field size of 14.19 starters, and the mean final win odds is 66.8 with a standard deviation of 96.35, reflecting the wide dispersion in ex ante market expectations.

\begin{table}[t!]
  \begin{center}
  \caption{Summary Statistics of Win-Odds Data}
  \label{tb:summary_statistics_horse_race}
  \subfloat[Race level]{
  
\begin{tabular}[t]{lrrrrr}
\toprule
  & N & Mean & SD & Min. & Max.\\
\midrule
Num Horses & 63372 & 14.19 & 2.61 & 4.00 & 18.00\\
Distance & 63372 & 1663.91 & 444.68 & 1000.00 & 4260.00\\
Num. of Wagers (Mil) & 63351 & 0.40 & 0.73 & 0.01 & 42.63\\
Racetrack Quality & 63372 & 3.57 & 0.80 & 1.00 & 4.00\\
\bottomrule
\end{tabular}

  }\\
  \subfloat[Horse-race level]{
  
\begin{tabular}[t]{lrrrrr}
\toprule
  & N & Mean & SD & Min. & Max.\\
\midrule
Horse Age & 895090 & 3.65 & 1.35 & 2.00 & 13.00\\
Realized Rank & 895090 & 7.81 & 4.43 & 1.00 & 18.00\\
Final Odds & 895090 & 66.80 & 96.35 & 1.10 & 999.90\\
Realized Finish Time & 895090 & 102.05 & 29.92 & 53.80 & 326.40\\
\bottomrule
\end{tabular}

  }
  \end{center}
  \footnotesize
  Note: The win-odds data cover all centrally administered races in Japan from 2004 to 2023.
\end{table}

\paragraph{Objective Winning Probability and Final Odds}

\begin{figure}[t!]
\centering
\caption{Winning Probability versus Final Odds}
         \centering
        \includegraphics[width=0.8\textwidth]{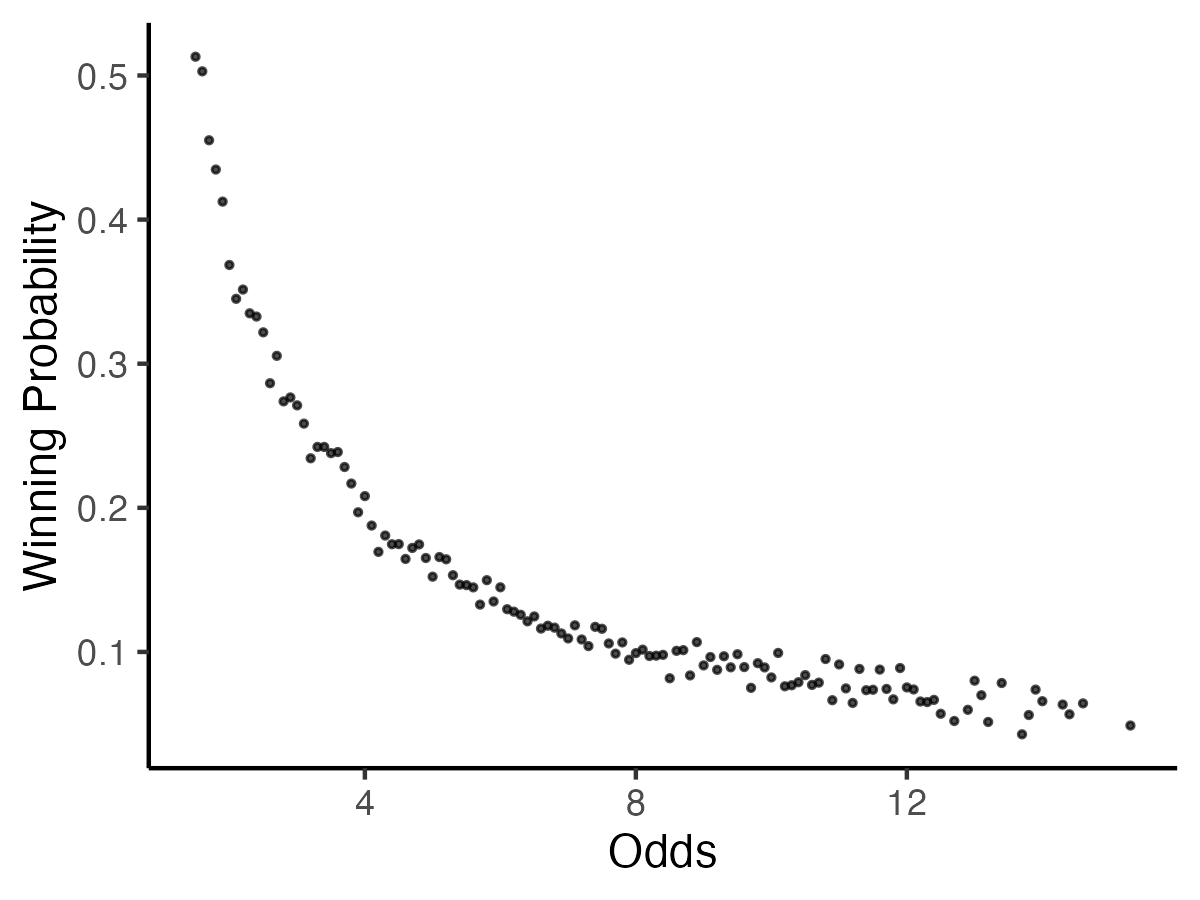}
        
       \label{fg:plot_winning_probability}

\vspace{1mm}
 \raggedright
      \begin{minilinespace}
  \begin{footnotesize}

    \textit{Note}: The figure plots empirical winning probabilities against final win odds. Horses with lower final odds (favorites) exhibit higher realized win rates, while those with higher odds (longshots) win far less often.
  \end{footnotesize}
  \end{minilinespace}
 \label{fg:plot_winning_probability_expected_return} 
\end{figure}

Figure~\ref{fg:plot_winning_probability_expected_return} depicts the relationship between final odds and (ex-post) winning probabilities in the JRA betting market. It illustrates a clear inverse relationship between final odds and realized win probabilities: horses with lower odds (i.e., favorites) exhibit substantially higher empirical win rates, while those with higher odds (i.e., longshots) show markedly lower chances of winning.

\newpage

\section{Portfolio Returns}\label{sec:portfolio_return}
 \renewcommand\theequation{\thesection.\arabic{equation}}
 \renewcommand\thefigure{\thesection.\arabic{figure}} 
  \renewcommand\thetable{\thesection.\arabic{table}} 
  \setcounter{equation}{0}
 \setcounter{figure}{0}
  \setcounter{table}{0}

This appendix describes how we recover market shares from observed odds, infer the portfolio of wagers placed during the final five minutes, and construct the corresponding realized returns.

For horse \(i\) in race \(r\), let \(R_{i,r,-\tau}\) denote the interim odds observed \(\tau\) minutes before post time, and let \(s_{i,r,-\tau}\) denote the corresponding share of cumulative wagers. Under the JRA's 20\% takeout rate, the two are related by
$R_{i,r,-\tau}
=0.8/s_{i,r,-\tau}$. Hence, the interim market share is recovered from the observed odds as
\[
s_{i,r,-\tau}
=
\frac{0.8}{R_{i,r,-\tau}}.
\]
Similarly, letting \(R^\ast_{i,r}\) denote the final odds, the final market share is
\[
s_{i,r,0}
=
\frac{0.8}{R^\ast_{i,r}}.
\]

Because posted odds are rounded, we normalize the recovered shares within each race and time point as \(\widetilde{s}_{i,r,-\tau}=s_{i,r,-\tau}/\sum_j s_{j,r,-\tau}\) and \(\widetilde{s}_{i,r,0}=s_{i,r,0}/\sum_j s_{j,r,0}\). For notational simplicity, the remainder of this appendix uses \(s_{i,r,-\tau}\) and \(s_{i,r,0}\) to denote these normalized shares.

Let \(B_{r,-5}\) and \(B_{r,0}\) denote cumulative betting volume five minutes before post time and at market closure. The fraction of the final pool wagered during the final five minutes in race $r$ is
\[
L_r
=
\frac{B_{r,0}-B_{r,-5}}{B_{r,0}}.
\]

Let \(\lambda^{late}_{i,r}\) denote the share of final-five-minute wagers allocated to horse \(i\) in race $r$. The final market share is then
\[
s_{i,r,0}
=
(1-L_r)s_{i,r,-5}
+
L_r\lambda^{late}_{i,r}.
\]
Solving for the implied late-betting share gives
\[
\lambda^{late}_{i,r}
=
\frac{
s_{i,r,0}
-
(1-L_r)s_{i,r,-5}
}{
L_r
}.
\]

We use these shares to construct the realized gross return on the late-betting portfolio in race $r$:
\[
R^{late}_r
=
\sum_i
\lambda^{late}_{i,r}
\mathbf{1}_{\{win_{i,r}=1\}}
R^\ast_{i,r}.
\]
For comparison, the portfolio represented by market shares five minutes before post time has realized gross return
\[
R^{early}_r
=
\sum_i
s_{i,r,-5}
\mathbf{1}_{\{win_{i,r}=1\}}
R^\ast_{i,r}.
\]
Because only one horse wins each race, each portfolio return equals the portfolio weight assigned to the winning horse multiplied by its final payout. These are aggregate, hypothetical portfolio returns rather than returns observed for individual bettors, whose betting records are unavailable.

\clearpage
\section{Robustness Analysis}
 \renewcommand\theequation{\thesection.\arabic{equation}}
 \renewcommand\thefigure{\thesection.\arabic{figure}} 
  \renewcommand\thetable{\thesection.\arabic{table}} 
  \setcounter{equation}{0}
 \setcounter{figure}{0}
  \setcounter{table}{0}

\subsection{Robustness to Control Specifications}\label{sec:robustness_fe}

This subsection examines the robustness of the baseline regression results to alternative control specifications. The main specification in Table~\ref{tab:main_results} includes a set of horse- and race-level controls, as well as race-year, racetrack-location, race-grade, horse, and jockey fixed effects. To assess whether the results are driven by these controls, we consider two alternative specifications.

First, Table~\ref{tb:estimation_results_odds_nocontrols} reports estimates from specifications without additional controls or fixed effects. The results remain qualitatively similar to the baseline estimates. In particular, the coefficient on the final-five-minute odds change remains negative and statistically significant, indicating that the association between late odds declines and higher realized returns is not driven by the inclusion of the baseline control variables.

Second, Table~\ref{tb:estimation_results_odds_race_fe} reports estimates from specifications that include race fixed effects. This specification absorbs all race-level heterogeneity common to horses competing in the same race, including unobserved features such as race-specific information, field composition, and aggregate betting conditions. The main coefficient of interest remains negative and statistically significant, confirming that the relationship between final-five-minute odds changes and realized returns persists even when identification comes from within-race variation across horses.

Overall, these robustness checks show that the main findings are not sensitive to the particular choice of control variables or fixed effects. The evidence continues to indicate that, conditional on final odds, horses whose odds decline near post time tend to yield higher realized returns.

  \begin{table}[t!]
  \centering
  \footnotesize
  \caption{Estimation Results without Controls}
  \label{tb:estimation_results_odds_nocontrols}
  \begin{center}
  
\begin{tabular}[t]{lcccccc}
\toprule
  & (1) & (2) & (3) & (4) & (5) & (6)\\
\midrule
Constant & 0.8338 & 0.8452 & 0.8443 & 0.8435 & 0.8423 & 0.8333\\
 & (0.0074) & (0.0080) & (0.0082) & (0.0082) & (0.0082) & (0.0083)\\
$R_{i}^{*}$ & -0.0016 & -0.0012 & -0.0012 & -0.0013 & -0.0014 & -0.0012\\
 & (0.0001) & (0.0001) & (0.0001) & (0.0001) & (0.0001) & (0.0001)\\
$\frac{\Delta R_{i,[-5, 0]}}{R_{i,-5}}$ &  & -0.3559 &  &  &  & -0.4416\\
 &  & (0.0381) &  &  &  & (0.0407)\\
$\frac{\Delta R_{i,[-10, 0]}}{R_{i,-10}}$ &  &  & -0.1674 &  &  & \\
 &  &  & (0.0250) &  &  & \\
$\frac{\Delta R_{i,[-15, 0]}}{R_{i,-15}}$ &  &  &  & -0.1052 &  & \\
 &  &  &  & (0.0210) &  & \\
$\frac{\Delta R_{i,[-20, 0]}}{R_{i,-20}}$ &  &  &  &  & -0.0702 & \\
 &  &  &  &  & (0.0176) & \\
$\frac{\Delta R_{i,[-10, -5]}}{R_{i,-10}}$ &  &  &  &  &  & 0.1526\\
 &  &  &  &  &  & (0.0598)\\
$\frac{\Delta R_{i,[-15, -10]}}{R_{i,-15}}$ &  &  &  &  &  & 0.3653\\
 &  &  &  &  &  & (0.0889)\\
$\frac{\Delta R_{i,[-20, -15]}}{R_{i,-20}}$ &  &  &  &  &  & 0.2211\\
 &  &  &  &  &  & (0.0777)\\
$R_{i}^{*} \times \frac{\Delta R_{i,[-5, 0]}}{R_{i,-5}}$ &  & 0.0002 &  &  &  & 0.0006\\
 &  & (0.0003) &  &  &  & (0.0003)\\
$R_{i}^{*} \times \frac{\Delta R_{i,[-10, 0]}}{R_{i,-10}}$ &  &  & 0.0000 &  &  & \\
 &  &  & (0.0002) &  &  & \\
$R_{i}^{*} \times \frac{\Delta R_{i,[-15, 0]}}{R_{i,-15}}$ &  &  &  & 0.0001 &  & \\
 &  &  &  & (0.0001) &  & \\
$R_{i}^{*} \times \frac{\Delta R_{i,[-20, 0]}}{R_{i,-20}}$ &  &  &  &  & 0.0001 & \\
 &  &  &  &  & (0.0001) & \\
$R_{i}^{*} \times \frac{\Delta R_{i,[-10, -5]}}{R_{i,-10}}$ &  &  &  &  &  & -0.0009\\
 &  &  &  &  &  & (0.0004)\\
$R_{i}^{*} \times \frac{\Delta R_{i,[-15, -10]}}{R_{i,-15}}$ &  &  &  &  &  & -0.0007\\
 &  &  &  &  &  & (0.0006)\\
$R_{i}^{*} \times \frac{\Delta R_{i,[-20, -15]}}{R_{i,-20}}$ &  &  &  &  &  & -0.0007\\
 &  &  &  &  &  & (0.0006)\\
\midrule
Num.Obs. & 894127 & 894127 & 894127 & 894127 & 894127 & 894127\\
R2 & 0.001 & 0.001 & 0.001 & 0.001 & 0.001 & 0.001\\
R2 Adj. & 0.001 & 0.001 & 0.001 & 0.001 & 0.001 & 0.001\\
\bottomrule
\end{tabular}

  \end{center}
     \begin{tablenotes}[flushleft]
\footnotesize
\item \textit{Notes:} This table reports OLS estimates of the coefficients from the regression:  
\[
 \mathbf{1}_{\{win_{i}=1\}}R_{i}^{\ast} = \alpha + \beta R_{i}^{\ast} + \delta \,  \text{OddsChange}_{i} + \gamma \,  R_{i}^{\ast} \times \text{OddsChange}_{i} + \varepsilon_{i}.
\]  
The variable \( \Delta R_{i,[-\tau,0]}/R_{i,-\tau} \equiv (R_{i}^{\ast} - R_{i,-\tau}) / R_{i,-\tau} \) represents the rate of change in odds over the final \( \tau \) minutes before post time.  
    \end{tablenotes}
\end{table}

\begin{table}[t!]
  
  \footnotesize
  \caption{Estimation Results with Race Fixed Effects}
  \label{tb:estimation_results_odds_race_fe}
  \begin{center}
      
\begin{tabular}[t]{lcccccc}
\toprule
  & (1) & (2) & (3) & (4) & (5) & (6)\\
\midrule
$R_{i}^{*}$ & -0.00158 & -0.00093 & -0.00100 & -0.00121 & -0.00125 & -0.00096\\
 & (0.00010) & (0.00020) & (0.00022) & (0.00021) & (0.00021) & (0.00025)\\
$\frac{\Delta R_{i,[-5, 0]}}{R_{i,-5}}$ &  & -0.35756 &  &  &  & -0.44458\\
 &  & (0.03946) &  &  &  & (0.03917)\\
$\frac{\Delta R_{i,[-10, 0]}}{R_{i,-10}}$ &  &  & -0.16842 &  &  & \\
 &  &  & (0.02844) &  &  & \\
$\frac{\Delta R_{i,[-15, 0]}}{R_{i,-15}}$ &  &  &  & -0.10831 &  & \\
 &  &  &  & (0.02453) &  & \\
$\frac{\Delta R_{i,[-20, 0]}}{R_{i,-20}}$ &  &  &  &  & -0.07297 & \\
 &  &  &  &  & (0.02048) & \\
$\frac{\Delta R_{i,[-10, -5]}}{R_{i,-10}}$ &  &  &  &  &  & 0.1602\\
 &  &  &  &  &  & (0.0651)\\
$\frac{\Delta R_{i,[-15, -10]}}{R_{i,-15}}$ &  &  &  &  &  & 0.3266\\
 &  &  &  &  &  & (0.0744)\\
$\frac{\Delta R_{i,[-20, -15]}}{R_{i,-20}}$ &  &  &  &  &  & 0.2157\\
 &  &  &  &  &  & (0.0658)\\
$R_{i}^{*} \times \frac{\Delta R_{i,[-5, 0]}}{R_{i,-5}}$ &  & -0.00030 &  &  &  & 0.00017\\
 &  & (0.00036) &  &  &  & (0.00036)\\
$R_{i}^{*} \times \frac{\Delta R_{i,[-10, 0]}}{R_{i,-10}}$ &  &  & -0.00023 &  &  & \\
 &  &  & (0.00029) &  &  & \\
$R_{i}^{*} \times \frac{\Delta R_{i,[-15, 0]}}{R_{i,-15}}$ &  &  &  & -0.00007 &  & \\
 &  &  &  & (0.00025) &  & \\
$R_{i}^{*} \times \frac{\Delta R_{i,[-20, 0]}}{R_{i,-20}}$ &  &  &  &  & -0.00008 & \\
 &  &  &  &  & (0.00020) & \\
$R_{i}^{*} \times \frac{\Delta R_{i,[-10, -5]}}{R_{i,-10}}$ &  &  &  &  &  & -0.00075\\
 &  &  &  &  &  & (0.00078)\\
$R_{i}^{*} \times \frac{\Delta R_{i,[-15, -10]}}{R_{i,-15}}$ &  &  &  &  &  & -0.00050\\
 &  &  &  &  &  & (0.00053)\\
$R_{i}^{*} \times \frac{\Delta R_{i,[-20, -15]}}{R_{i,-20}}$ &  &  &  &  &  & -0.00115\\
 &  &  &  &  &  & (0.00053)\\
\midrule
Num.Obs. & 894127 & 894127 & 894127 & 894127 & 894127 & 894127\\
R2 & 0.054 & 0.054 & 0.054 & 0.054 & 0.054 & 0.054\\
R2 Adj. & -0.018 & -0.018 & -0.018 & -0.018 & -0.018 & -0.018\\
\bottomrule
\end{tabular}

  \end{center}
  
     \begin{tablenotes}[flushleft]
\footnotesize
\item \textit{Notes:} This table reports OLS estimates of the coefficients from the regression:  
\[
\mathbf{1}_{\{win_{i}=1\}}R_{i}^{\ast} = \alpha_{j(i)} +  \beta R_{i}^{\ast} + \delta \,  \text{OddsChange}_{i} + \gamma \,  R_{i}^{\ast} \times \text{OddsChange}_{i} + \text{Race Fixed Effect} + \varepsilon_{i}.
\]  
The variable \( \Delta R_{i,[-\tau,0]}/R_{i,-\tau} \equiv (R_{i}^{\ast} - R_{i,-\tau}) / R_{i,-\tau} \) denotes the rate of change in odds over the final \( \tau \) minutes before post time. 
    \end{tablenotes}
\end{table}

\clearpage
\subsection{Decomposing Last-Five-Minute Odds Changes}
\label{appen_last_five_minute_decomposition}

The main analysis focuses on odds changes over the final five minutes before post time. This window is empirically important because, as documented in Section~\ref{Sec:Data}, a large fraction of wagering activity is concentrated immediately before post time. At the same time, the five-minute window is still relatively coarse. As described in Supplementary Appendix \ref{sec:appendix_data}, in the JRA betting system, betting remains open after the five-minute mark, and substantial wagering activity may occur within this interval. Thus, the change from five-minute odds to final odds may combine information arriving at different points within the final betting window. In principle, observing interim odds continuously during these final minutes would allow us to examine more precisely when late information is incorporated into prices.

A practical difficulty is that odds updates after the five-minute mark are not uniform across races. In some races, the five-minute odds are the last available interim odds before final odds are posted. In others, odds are updated again, but the timing of these updates varies across races. Among post-five-minute updates, the two-minute mark is the most frequently observed update time. We therefore focus on the subsample of races for which interim odds are observed at two minutes before post time. This subsample accounts for roughly two-thirds of the full sample.

Using this subsample, we decompose the last-five-minute odds change into two components. The first captures the change from five to two minutes before post time, while the second captures the change from two minutes before post time to the final odds. Specifically, we define the five-to-two-minute odds change as
\(\Delta R_{i,[-5,-2]}/R_{i,-5} \equiv (R_{i,-2}-R_{i,-5})/R_{i,-5}\),
and the two-minute-to-final odds change as
\(\Delta R_{i,[-2,0]}/R_{i,-2} \equiv (R_i^{\ast}-R_{i,-2})/R_{i,-2}\). We then estimate regressions that include both components of the last-five-minute odds change, together with their interactions with final odds.

The results are reported in Table~\ref{tb:estimation_results_odds_1min_decomposition_5v2}. We focus on Column~(3), which includes the full set of controls and fixed effects. The coefficient on 
\(\Delta R_{i,[-5,-2]}/R_{i,-5}\) is \(-0.350\), while the coefficient on 
\(\Delta R_{i,[-2,0]}/R_{i,-2}\) is \(-1.465\). These magnitudes imply that a 10\% decline in odds between five and two minutes before post time is associated with realized gross returns that are about 3.5 percentage points higher on average. By contrast, a 10\% decline between two minutes before post time and the final odds is associated with returns that are about 14.7 percentage points higher.

Thus, both components of the last-five-minute odds change are negatively associated with realized returns, but the magnitude is substantially larger for the very last segment. This pattern suggests that odds revisions closer to market closure contain particularly strong information about expected returns, consistent with the importance of last-minute wagering activity.

\begin{table}[t!]
  \footnotesize
  \caption{Two-Interval Decomposition of Last-Five-Minute Odds Changes}
  \label{tb:estimation_results_odds_1min_decomposition_5v2}
  \begin{center}
      
\begin{tabular}[t]{lccc}
\toprule
  & (1) No FE & (2) Race FE & (3) Rich FE\\
\midrule
Constant & 0.84638 &  & \\
 & (0.00990) &  & \\
$R_{i}^{*}$ & -0.00112 & -0.00089 & -0.00092\\
 & (0.00014) & (0.00023) & (0.00023)\\
$\frac{\Delta R_{i,[-2, 0]}}{R_{i,-2}}$ & -1.61784 & -1.84559 & -1.46526\\
 & (1.07869) & (0.71756) & (0.68883)\\
$\frac{\Delta R_{i,[-5, -2]}}{R_{i,-5}}$ & -0.36353 & -0.35874 & -0.35023\\
 & (0.04579) & (0.04473) & (0.04360)\\
$R_{i}^{*} \times \frac{\Delta R_{i,[-2, 0]}}{R_{i,-2}}$ & -0.00153 & 0.00232 & -0.00327\\
 & (0.01070) & (0.00644) & (0.00660)\\
$R_{i}^{*} \times \frac{\Delta R_{i,[-5, -2]}}{R_{i,-5}}$ & 0.00019 & -0.00030 & 0.00005\\
 & (0.00031) & (0.00039) & (0.00036)\\
\midrule
Num.Obs. & 611846 & 611846 & 611846\\
R2 & 0.001 & 0.057 & 0.002\\
R2 Adj. & 0.001 & -0.018 & 0.001\\
FE: Race &  & X & \\
FE: Race-Year &  &  & X\\
FE: Race-Location &  &  & X\\
FE: Race-Grade &  &  & X\\
FE: Jockey &  &  & X\\
Horse Characteristics &  &  & X\\
\bottomrule
\end{tabular}

  \end{center}

       \begin{tablenotes}[flushleft]
\footnotesize
\item \textit{Notes:} The dependent variable is $\mathbf{1}_{\{win_{i}=1\}}R_{i}^{\ast}$. Columns~(1)--(3) correspond to No~FE, Race~FE, and Rich~FE specifications. Column~(2) saturates race-level variation: race fixed effects absorb all regressors that are constant within a race. Column~(3) uses a richer fixed-effects structure (race-year, racetrack location, race grade, horse, and jockey FE, together with horse and race controls) that does not saturate the race level, so race-level regressors and their interactions remain identified. Standard errors are clustered at the race level in Columns~(2)--(3).
    \end{tablenotes}
\end{table}

\clearpage
\subsection{Ex-ante Return Predictability}\label{App:ExAntePredict}

The results in Section~\ref{Sec:Results} document a significant association between realized returns, \(\mathbf{1}_{\{win_i=1\}}R_i^{\ast}\), and changes in odds from interim to final values in the last few minutes before the race. For clarity, these results should not be taken as evidence of \emph{ex-ante} return predictability. Rather, the correlations we document reflect \emph{ex-post} associations, because bettors cannot react after observing the final odds or place contingent orders based on them.

\begin{figure}[t!]
\caption{Expected Return versus Odds Changes: Interim Odds on the Horizontal Axis}
  \label{fg:plot_expected_return_intermediate_odds_min_before_interim_xaxis} 
  \begin{center}
  \subfloat[-5 to 0 (final)]{\includegraphics[width = 0.375\textwidth]{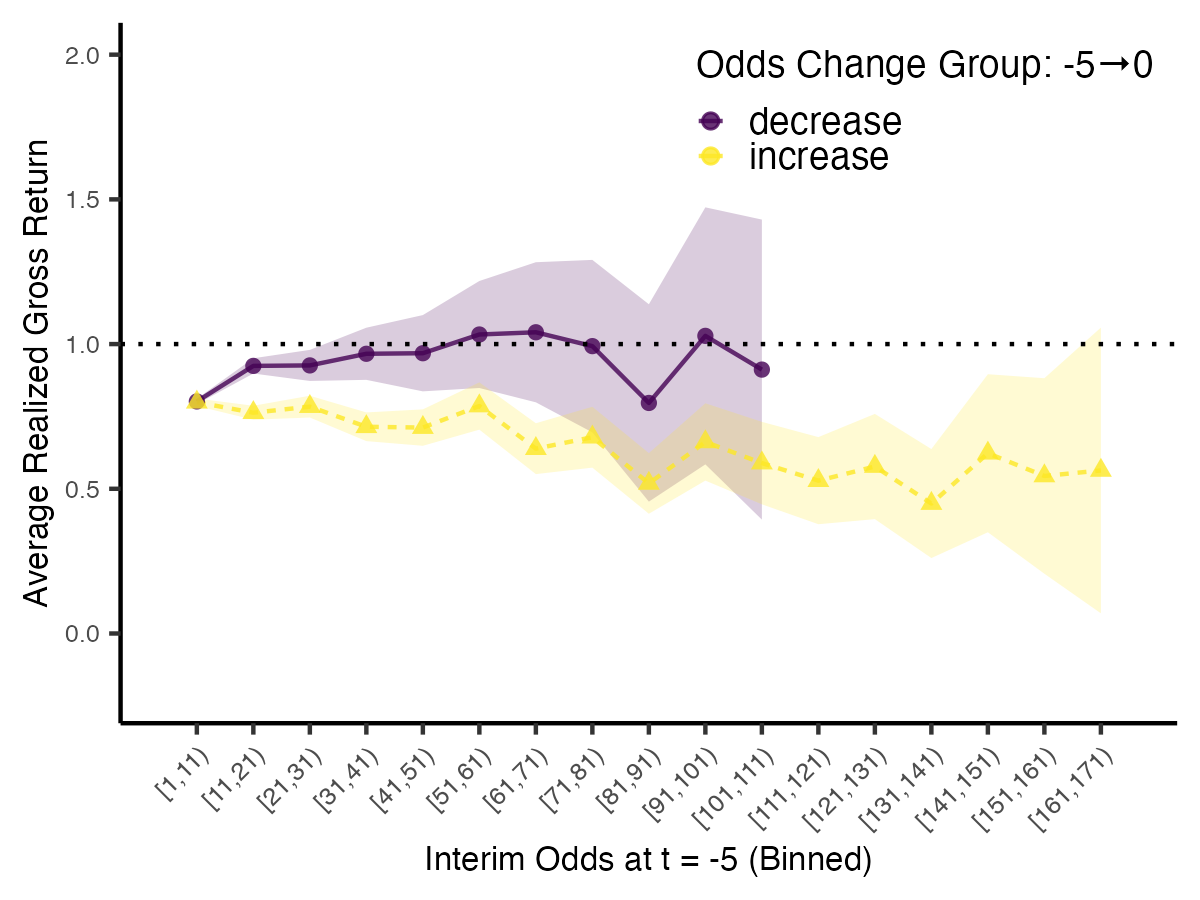}}
  \qquad\qquad
  \subfloat[-10 to -5]{\includegraphics[width = 0.375\textwidth]{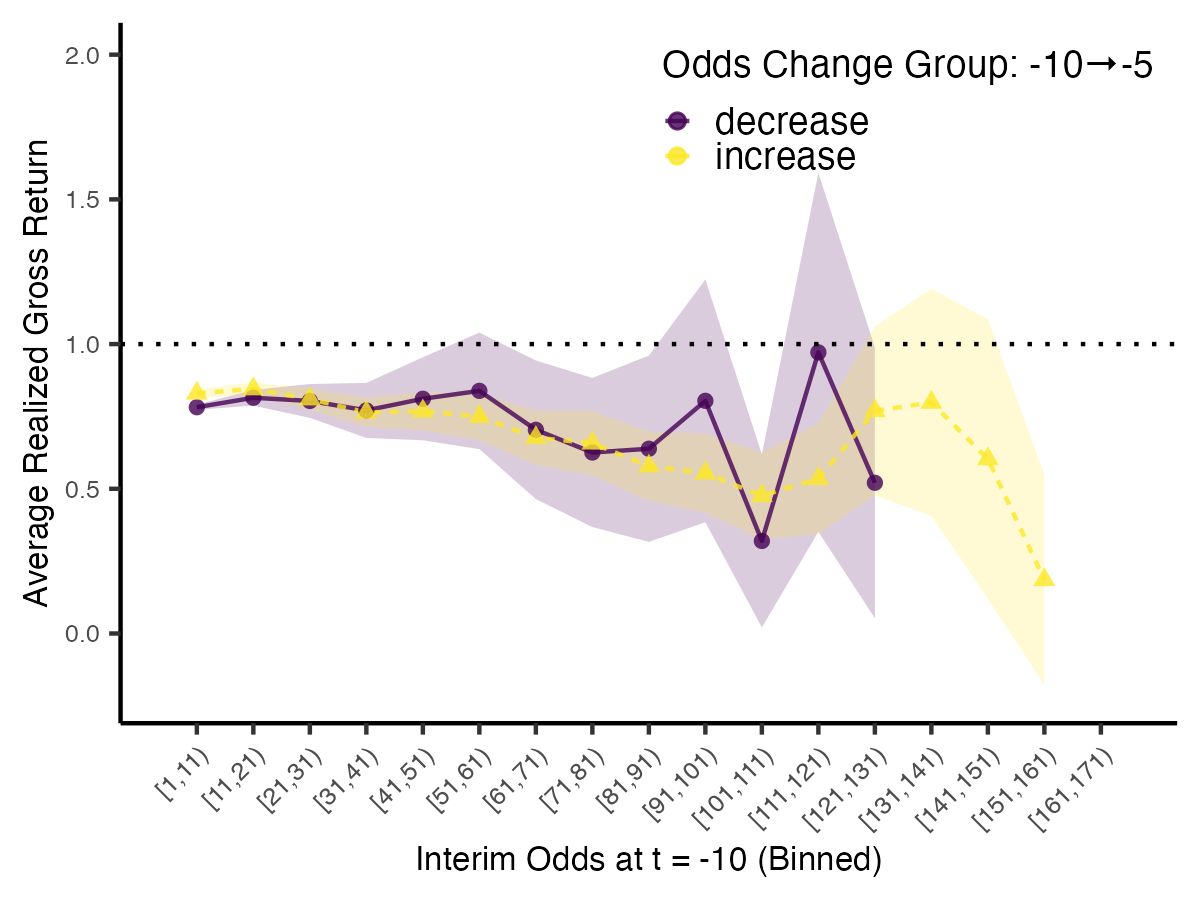}}\\
  \subfloat[-15 to -10]{\includegraphics[width = 0.375\textwidth]{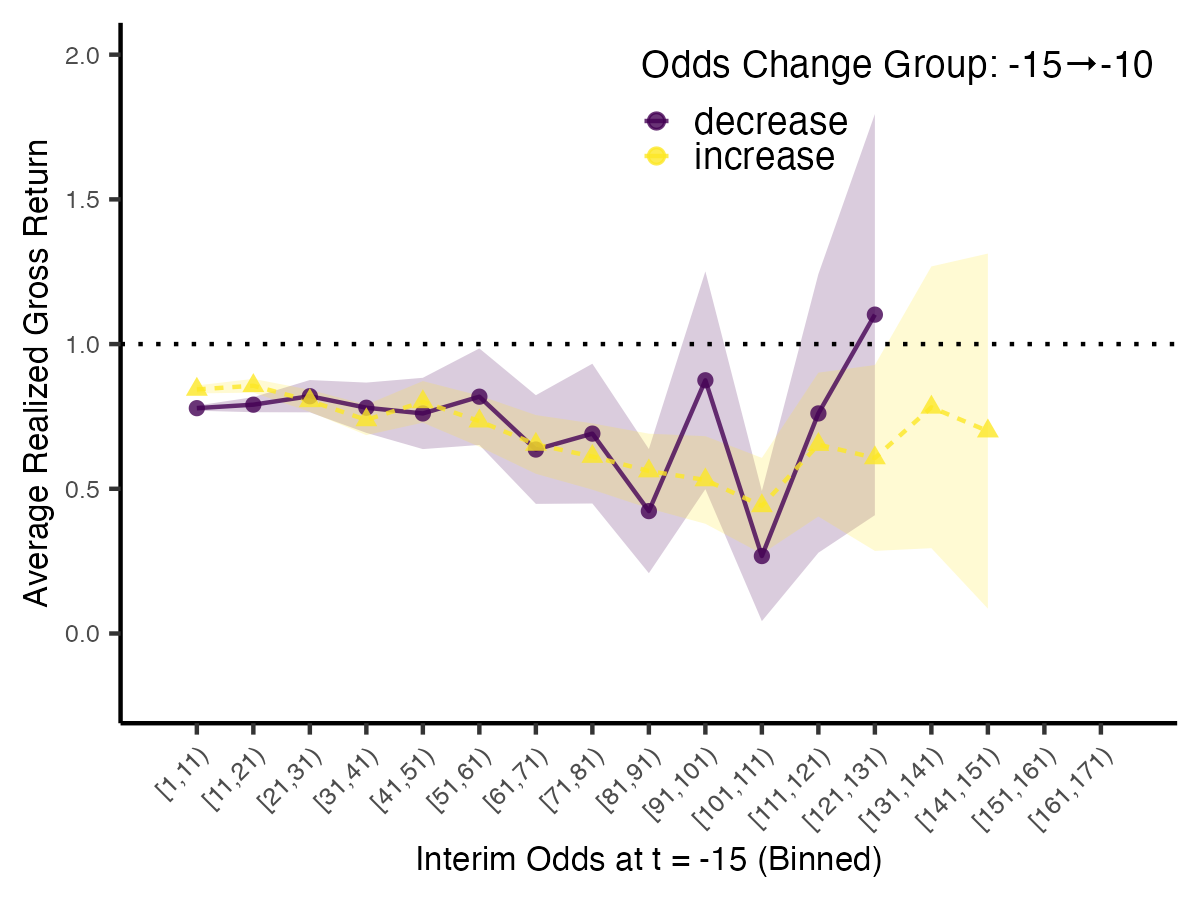}}
  \qquad\qquad
  \subfloat[-20 to -15]{\includegraphics[width = 0.375\textwidth]{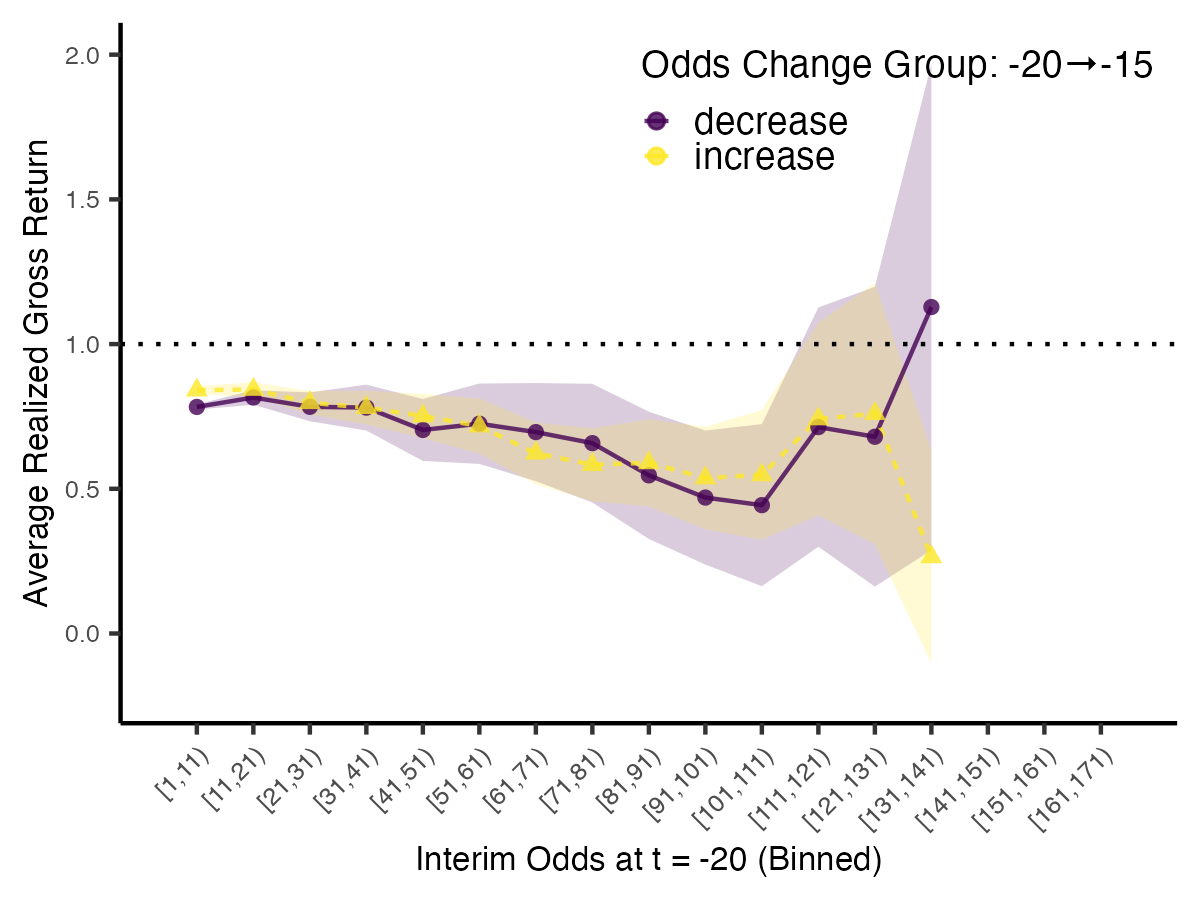}}\\
  \subfloat[-25 to -20]{\includegraphics[width = 0.375\textwidth]{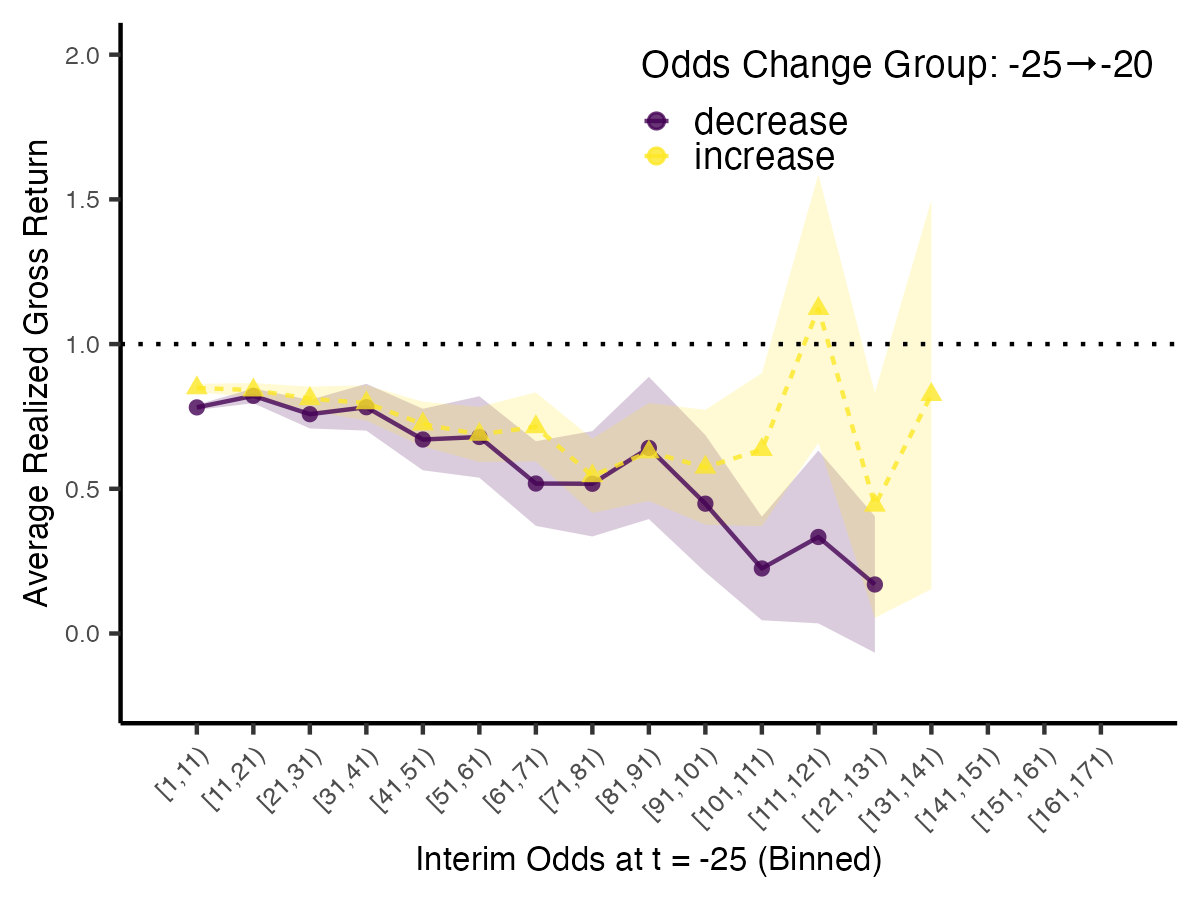}}
  \qquad\qquad
  \subfloat[-30 to -25]{\includegraphics[width = 0.375\textwidth]{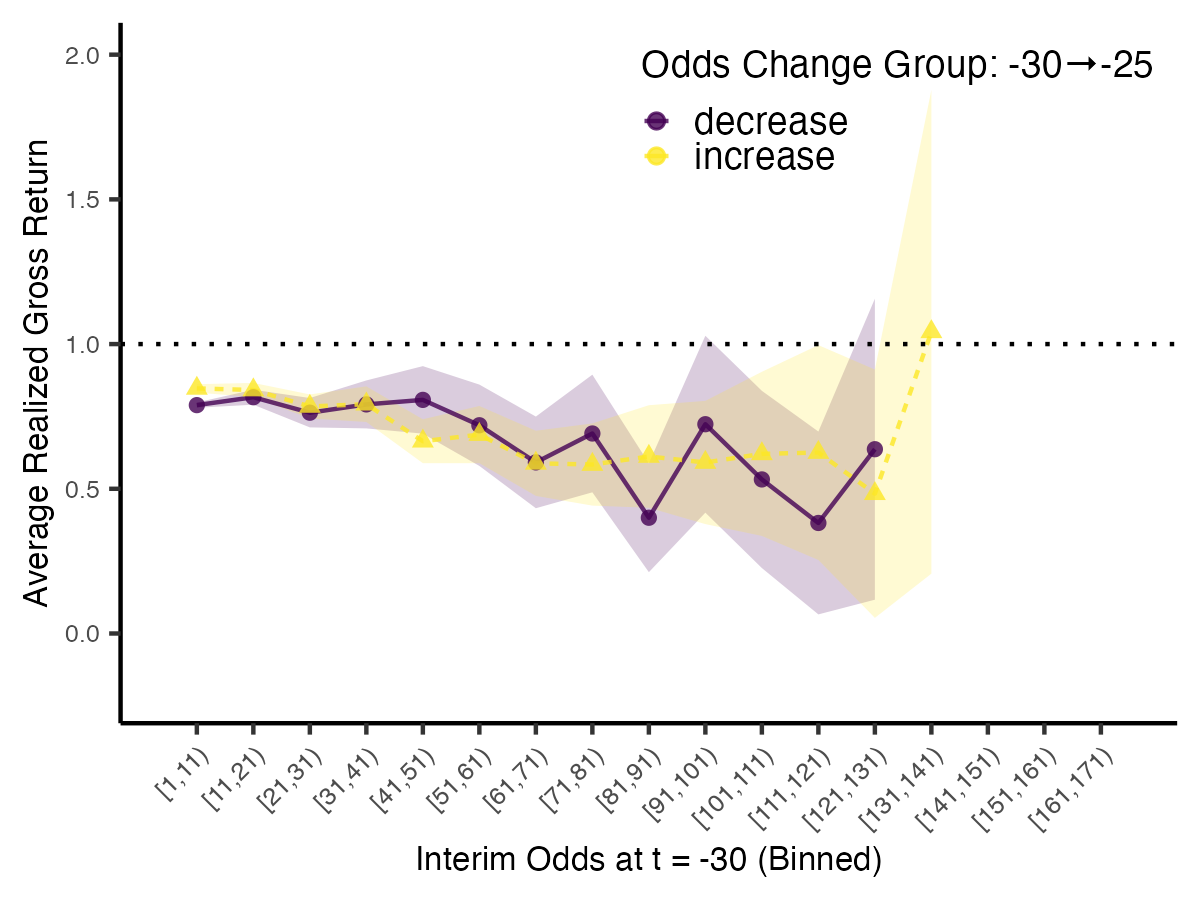}}

  \end{center}

\begin{tablenotes}[flushleft]
\footnotesize
\item \textit{Notes:}  The horizontal axis represents the latest interim odds available at the start of each interval, grouped into bins of width 10. Panels (a)--(f) compare average realized returns for horses whose odds increased versus decreased during specified five-minute intervals before post time. Panel (a) covers the final interval, with the horizontal axis based on odds observed five minutes before post time. Panels (b)--(f) examine earlier intervals, with the horizontal axis based on odds observed at the start of each corresponding interval. Within each panel, the two lines plot average realized returns for horses with increasing and decreasing odds during the corresponding interval. Shaded areas denote 95\% confidence intervals. Bins with fewer than 1,000 observations are excluded separately by group.
\end{tablenotes}
\end{figure}

To illustrate this point, Figure~\ref{fg:plot_expected_return_intermediate_odds_min_before_interim_xaxis} presents an alternative version of Figure~\ref{fg:plot_expected_return_intermediate_odds_min_before_5_to_0}. The key difference is the variable on the horizontal axis. While Figure~\ref{fg:plot_expected_return_intermediate_odds_min_before_5_to_0} groups horses by final odds, Figure~\ref{fg:plot_expected_return_intermediate_odds_min_before_interim_xaxis} groups them by the latest interim odds observed at the start of each interval. Thus, Panel~(a), which covers the final five-minute interval, uses odds observed five minutes before post time rather than final odds. Panels~(b)--(f) similarly use the interim odds observed at the start of each corresponding five-minute interval.

This distinction matters for assessing whether the patterns can be interpreted as \emph{ex-ante} information. In Panel~(a), bettors can observe the five-minute odds shown on the horizontal axis, but they cannot observe the subsequent change from five-minute to final odds before making their betting decisions. Thus, the increase/decrease classification in Panel~(a) is still based on information realized only after the five-minute mark. By contrast, in Panels~(b)--(f), both the interim odds level and the odds change over the corresponding interval are observable before the final minutes of the betting window. These panels therefore provide a closer approximation to information that could have been available to bettors before market closure.

Overall, Figure~\ref{fg:plot_expected_return_intermediate_odds_min_before_interim_xaxis} indicates that the strong return differences associated with odds declines are concentrated in the final-five-minute interval. In Panel~(a), horses whose odds decline between five minutes before post time and the final odds tend to earn higher realized returns, consistent with the main results. However, this classification relies on odds changes that are not observable five minutes before post time, so the association in Panel~(a) should be interpreted as an \emph{ex-post} pattern. By contrast, in Panels~(b)--(f), where both interim odds levels and odds changes are observable before the final minutes of betting, the differences are much weaker. These patterns provide limited support for exploitable \emph{ex-ante} return predictability based on earlier odds movements.

\clearpage

\section{Additional Empirical Analysis}\label{sec:additional_empirical}
 \renewcommand\theequation{\thesection.\arabic{equation}}
 \renewcommand\thefigure{\thesection.\arabic{figure}} 
  \renewcommand\thetable{\thesection.\arabic{table}} 
  \setcounter{equation}{0}
 \setcounter{figure}{0}
  \setcounter{table}{0}

\subsection{Asymmetry in Odds Movements}\label{sec:robustness_asymmetry}

We decompose the last-five-minute odds change into positive and negative components:
\[
\Delta_i^{+}
=
\max\left\{
\frac{\Delta R_{i,[-5,0]}}{R_{i,-5}},\,0
\right\},
\qquad
\Delta_i^{-}
=
\min\left\{
\frac{\Delta R_{i,[-5,0]}}{R_{i,-5}},\,0
\right\}.
\]
Here, \(\Delta_i^{+}\ge 0\) captures odds increases, while \(\Delta_i^{-}\le 0\) captures odds declines. We then estimate
\[
\mathbf{1}_{\{win_i=1\}}R_i^{\ast}
= \alpha
+\beta R_i^{\ast}
+\delta^{+}\Delta_i^{+}
+\delta^{-}\Delta_i^{-} +\gamma^{+} R_i^{\ast}\times \Delta_i^{+}
+\gamma^{-} R_i^{\ast}\times \Delta_i^{-}
+Z_i'\zeta
+\varepsilon_i.
\]

In this specification, the effect of a one-percent odds increase on the expected gross return, evaluated at final odds \(R_i^\ast\), is
\[
0.01\left(\delta^{+}+\gamma^{+}R_i^\ast\right).
\]
By contrast, the effect of a one-percent odds decline is
\[
-0.01\left(\delta^{-}+\gamma^{-}R_i^\ast\right),
\]
because a one-percent odds decline corresponds to \(\Delta_i^{-}=-0.01\). Thus, if \(\delta^{+}=\delta^{-}\) and \(\gamma^{+}=\gamma^{-}\), odds increases and odds declines have symmetric effects in magnitude but opposite signs.

Table~\ref{tb:estimation_results_odds_asymmetric_change} provides regression evidence corresponding to Panel~(a) of Figure~\ref{fg:plot_expected_return_intermediate_odds_min_before_5_to_0}. Both positive and negative odds movements are associated with subsequent returns in the expected direction: odds increases predict lower returns, whereas odds decreases predict higher returns. At \(R_i^\ast=0\), the coefficients on \(\Delta_i^+\) and \(\Delta_i^-\) do not indicate strong asymmetry. However, the interaction terms with final odds suggest that the predictive effect is stronger when odds decline. For example, using Column~(3), at final odds \(R_i^\ast=20\), a one-percent odds increase changes the expected return by approximately
\[
0.01(-0.32185+0.00014\times 20) \simeq -0.0032,
\]
whereas a one-percent odds decline changes the expected return by approximately
\[
-0.01(-0.26636-0.00667\times 20) \simeq 0.0040.
\]
Thus, at higher odds, odds declines have a larger positive predictive effect than the negative effect associated with odds increases. This asymmetry is consistent with Panel~(a) of Figure~\ref{fg:plot_expected_return_intermediate_odds_min_before_5_to_0}, which shows that horses whose odds decline in the final five minutes exhibit higher realized returns than those whose odds increase.

\begin{table}[h!]
  \footnotesize
  \caption{Asymmetry in Odds Movements: Positive vs.\ Negative Changes}
  \label{tb:estimation_results_odds_asymmetric_change}
  \begin{center}
      
\begin{tabular}[t]{lccc}
\toprule
  & (1) & (2) & (3)\\
\midrule
$R_{i}^{*}$ & -0.00122 & -0.00103 & -0.00103\\
 & (0.00012) & (0.00021) & (0.00020)\\
$\Delta^{+}$ & -0.33277 & -0.39514 & -0.32185\\
 & (0.05258) & (0.06232) & (0.05958)\\
$\Delta^{-}$ & -0.31049 & -0.15659 & -0.26636\\
 & (0.11459) & (0.09101) & (0.08316)\\
$R_{i}^{*} \times \Delta^{+}$ & 0.00025 & -0.00004 & 0.00014\\
 & (0.00029) & (0.00039) & (0.00036)\\
$R_{i}^{*} \times \Delta^{-}$ & -0.00609 & -0.00745 & -0.00667\\
 & (0.00290) & (0.00405) & (0.00401)\\
\midrule
Wald $\chi^2$: $\Delta^{+} = \Delta^{-}$ & 0.025 & 3.435 & 0.226\\
$p$-value & {}[0.8749] & {}[0.0638] & {}[0.6343]\\
Wald $\chi^2$: $R_i^{*} \times \Delta^{+} = R_i^{*} \times \Delta^{-}$ & 4.553 & 3.184 & 2.772\\
$p$-value & {}[0.0329] & {}[0.0743] & {}[0.0959]\\
Num.Obs. & 894127 & 894127 & 894127\\
R2 & 0.001 & 0.054 & 0.002\\
R2 Adj. & 0.001 & -0.018 & 0.001\\
FE: Race &  & X & \\
FE: Race-Year &  &  & X\\
FE: Race-Location &  &  & X\\
FE: Race-Grade &  &  & X\\
FE: Jockey &  &  & X\\
Horse Characteristics &  &  & X\\
\bottomrule
\end{tabular}

  \end{center}
     \begin{tablenotes}[flushleft]
\footnotesize
\item \textit{Notes:} 
\(\Delta_i^{+}=\max\{\Delta R_{i,[-5,0]}/R_{i,-5},0\}\) captures odds increases, while
\(\Delta_i^{-}=\min\{\Delta R_{i,[-5,0]}/R_{i,-5},0\}\le 0\) captures odds decreases.
A one-percent odds increase changes the expected gross return by
\(0.01(\delta^{+}+\gamma^{+}R_i^\ast)\), whereas a one-percent odds decline changes it by
\(-0.01(\delta^{-}+\gamma^{-}R_i^\ast)\).
Wald tests report tests of symmetric effects, \(\delta^{+}=\delta^{-}\) and \(\gamma^{+}=\gamma^{-}\).
The dependent variable is \(\mathbf{1}_{\{win_{i}=1\}}R_{i}^{\ast}\).
Standard errors are clustered at the race level in Columns~(2)--(3).
    \end{tablenotes}
\end{table}

\clearpage
\subsection{Path Dependence across Adjacent Five-Minute Windows}\label{appen_path_dependence}
The main analysis shows that odds movements during the final five minutes contain substantially more information about realized returns than movements observed earlier. This pattern is consistent with an information-based explanation in which return-relevant private information is incorporated into odds close to market closure. To examine this explanation further, we compare ex post returns across cases in which final-five-minute odds movements continue earlier betting trends and cases in which they overturn the direction established by earlier betting activity.

Specifically, we estimate the following extended regression model:
\begin{align}
\mathbf{1}_{\{win_i=1\}}R_i^{\ast}
=&\ \alpha+\beta R_i^{\ast}
+\delta_L^{+}\Delta_{i,L}^{+}
+\delta_L^{-}\Delta_{i,L}^{-}
+\delta_E^{+}\Delta_{i,E}^{+}
+\delta_E^{-}\Delta_{i,E}^{-} 
+\kappa_{+-}\Delta_{i,E}^{+}\Delta_{i,L}^{-}
+\kappa_{-+}\Delta_{i,E}^{-}\Delta_{i,L}^{+} \nonumber\\
&\quad
+\gamma_L^{+}R_i^{\ast}\Delta_{i,L}^{+}
+\gamma_L^{-}R_i^{\ast}\Delta_{i,L}^{-}
+\gamma_E^{+}R_i^{\ast}\Delta_{i,E}^{+}
+\gamma_E^{-}R_i^{\ast}\Delta_{i,E}^{-} \nonumber\\
&\quad
+\lambda_{+-}R_i^{\ast}\Delta_{i,E}^{+}\Delta_{i,L}^{-}
+\lambda_{-+}R_i^{\ast}\Delta_{i,E}^{-}\Delta_{i,L}^{+}
+Z_i'\zeta+\varepsilon_i.
\label{eq:extended_symmetry}
\end{align}
Here, $\Delta_{i,L}\equiv\Delta R_{i,[-5,0]}/R_{i,-5}$ denotes the odds-change rate during the final five minutes, while $\Delta_{i,E}\equiv\Delta R_{i,[-10,-5]}/R_{i,-10}$ denotes the odds-change rate during the preceding five-minute interval. For each $k\in\{L,E\}$, we decompose the odds change into its positive and negative components: $\Delta_{i,k}^{+}\equiv\max\{\Delta_{i,k},0\}$ and $\Delta_{i,k}^{-}\equiv\min\{\Delta_{i,k},0\}$. Thus, $\Delta_{i,k}^{+}$ captures odds increases, whereas $\Delta_{i,k}^{-}$ captures odds declines. The interaction $\Delta_{i,E}^{+}\Delta_{i,L}^{-}$ captures an earlier odds increase followed by a final-five-minute decline, while $\Delta_{i,E}^{-}\Delta_{i,L}^{+}$ captures an earlier odds decline followed by a final-five-minute increase. The corresponding interactions with $R_i^{\ast}$ allow these reversal patterns to vary with the final-odds level.

Table~\ref{tb:estimation_results_odds_extended_symmetry} reports the estimates. We focus on column~(3), which includes the rich set of controls and fixed effects. The estimates preserve the central asymmetry in the final betting window. Because $\Delta_{i,L}^{-}\leq0$, the negative coefficient on $\Delta_{i,L}^{-}$ implies that larger odds declines during the final five minutes are associated with higher realized returns. By contrast, the coefficient on $\Delta_{i,E}^{-}$ is positive, implying that odds declines during the preceding five-minute interval are associated with lower realized returns. Thus, the return pattern associated with odds declines is specific to the final betting window rather than a continuation of a general relationship between downward odds movements and realized returns.

\begin{table}[h!]
  \footnotesize
  \caption{Extended Symmetry: Last Five-Minute vs.\ Preceding Five-Minute Odds Changes}
  \label{tb:estimation_results_odds_extended_symmetry}
  \begin{center}
      
\begin{tabular}[t]{lccc}
\toprule
  & (1) No FE & (2) Race FE & (3) Rich FE\\
\midrule
$R_{i}^{*}$ & -0.00118 & -0.00099 & -0.00097\\
 & (0.00015) & (0.00028) & (0.00027)\\
$\Delta_{i,L}^{+}$ & -0.40383 & -0.48592 & -0.39211\\
 & (0.05955) & (0.06889) & (0.06411)\\
$\Delta_{i,L}^{-}$ & -0.25245 & -0.10806 & -0.20255\\
 & (0.13453) & (0.10982) & (0.09469)\\
$\Delta_{i,E}^{+}$ & 0.06020 & 0.05032 & 0.03949\\
 & (0.08927) & (0.13104) & (0.11468)\\
$\Delta_{i,E}^{-}$ & 0.58956 & 0.69399 & 0.64506\\
 & (0.15706) & (0.11028) & (0.09154)\\
$\Delta_{i,E}^{+}\,\Delta_{i,L}^{-}$ & -1.80961 & -1.48114 & -1.82867\\
 & (1.06935) & (1.26859) & (0.94924)\\
$\Delta_{i,E}^{-}\,\Delta_{i,L}^{+}$ & -1.53933 & -1.98080 & -1.53154\\
 & (0.63827) & (0.49587) & (0.44836)\\
$R_{i}^{*} \times \Delta_{i,L}^{+}$ & 0.00054 & 0.00027 & 0.00040\\
 & (0.00032) & (0.00042) & (0.00038)\\
$R_{i}^{*} \times \Delta_{i,L}^{-}$ & -0.00453 & -0.00625 & -0.00524\\
 & (0.00452) & (0.00554) & (0.00591)\\
$R_{i}^{*} \times \Delta_{i,E}^{+}$ & -0.00078 & -0.00071 & -0.00072\\
 & (0.00050) & (0.00107) & (0.00099)\\
$R_{i}^{*} \times \Delta_{i,E}^{-}$ & 0.00061 & 0.00133 & 0.00027\\
 & (0.00219) & (0.00199) & (0.00177)\\
$R_{i}^{*} \times \Delta_{i,E}^{+}\,\Delta_{i,L}^{-}$ & 0.00996 & 0.01067 & 0.01059\\
 & (0.01850) & (0.02775) & (0.02892)\\
$R_{i}^{*} \times \Delta_{i,E}^{-}\,\Delta_{i,L}^{+}$ & 0.00510 & 0.00564 & 0.00497\\
 & (0.00294) & (0.00204) & (0.00189)\\
\midrule
Num.Obs. & 894127 & 894127 & 894127\\
R2 & 0.001 & 0.054 & 0.002\\
R2 Adj. & 0.001 & -0.018 & 0.001\\
FE: Race &  & X & \\
FE: Race-Year &  &  & X\\
FE: Race-Location &  &  & X\\
FE: Race-Grade &  &  & X\\
FE: Jockey &  &  & X\\
Horse Characteristics &  &  & X\\
\bottomrule
\end{tabular}

  \end{center}
     \begin{tablenotes}[flushleft]
\footnotesize
\vspace{2mm}
\item \textit{Notes:} The dependent variable is \(\mathbf{1}_{\{win_i=1\}}R_i^{\ast}\). \(\Delta_{i,L}=\Delta R_{i,[-5,0]}/R_{i,-5}\) is the final-five-minute odds-change rate and \(\Delta_{i,E}=\Delta R_{i,[-10,-5]}/R_{i,-10}\) is the immediately preceding five-minute change; each is decomposed into a positive and a negative component as \(\Delta_{i,k}^{+}\equiv\max\{\Delta_{i,k},0\}\) and \(\Delta_{i,k}^{-}\equiv\min\{\Delta_{i,k},0\}\), \(k\in\{L,E\}\). The two off-diagonal cross-window terms \(\Delta_{i,E}^{+}\Delta_{i,L}^{-}\) and \(\Delta_{i,E}^{-}\Delta_{i,L}^{+}\) capture the ``reversal'' sign patterns, and the corresponding triple interactions \(R_i^{\ast}\Delta_{i,E}^{+}\Delta_{i,L}^{-}\) and \(R_i^{\ast}\Delta_{i,E}^{-}\Delta_{i,L}^{+}\) allow these reversal effects to vary with the final-odds level \(R_i^{\ast}\). Columns~(1)--(3) correspond to No~FE, Race~FE, and Rich~FE specifications, with the same definitions as in Table~\ref{tb:estimation_results_odds_1min_decomposition_5v2}. Standard errors are clustered at the race level in Columns~(2)--(3).
    \end{tablenotes}
\end{table}

The cross-window interactions allow the return association of final-five-minute odds movements to depend on whether they continue or reverse the preceding movement. Because these terms also interact with final odds, we illustrate their quantitative implications by comparing continuation and reversal paths of equal magnitude $d>0$, holding final odds fixed at $R_i^{\ast}=R$.

First, consider horses whose odds decline by $d$ during the final five minutes $[-5,0]$, so that $\Delta_{i,L}^{-}=-d$ and $\Delta_{i,L}^{+}=0$. Under reversal, odds increase by $d$ during the preceding interval $[-10,-5]$, so that $\Delta_{i,E}^{+}=d$ and $\Delta_{i,E}^{-}=0$. Under continuation, odds decline by $d$ over the same interval, so that $\Delta_{i,E}^{-}=-d$ and $\Delta_{i,E}^{+}=0$. From equation~\eqref{eq:extended_symmetry}, the difference in expected gross returns between the reversal and continuation paths is
\[
\begin{aligned}
&
\mathbb{E}\!\left[
\mathbf{1}_{\{win_i=1\}}R_i^{\ast}
\mid
R_i^{\ast}=R,\,
\Delta_{i,L}^{-}=-d,\,
\Delta_{i,E}^{+}=d
\right]
\\
&\quad-
\mathbb{E}\!\!\left[
\mathbf{1}_{\{win_i=1\}}R_i^{\ast}
\mid
R_i^{\ast}=R,\,
\Delta_{i,L}^{-}=-d,\,
\Delta_{i,E}^{-}=-d
\right]
\\
&=
d(\delta_E^{+}+\delta_E^{-})
+
Rd(\gamma_E^{+}+\gamma_E^{-})
-
d^2\kappa_{+-}
-
Rd^2\lambda_{+-}.
\end{aligned}
\]

Evaluating this expression at $R=20$ and $d=0.1$ using the point estimates from the rich-fixed-effects specification (column~(3) of Table~\ref{tb:estimation_results_odds_extended_symmetry}) gives
\[
\begin{aligned}
&0.1(0.03949+0.64506)
+20(0.1)(-0.00072+0.00027)\\
&\qquad
-(0.1)^2(-1.82867)
-20(0.1)^2(0.01059)
=0.0837.
\end{aligned}
\]
Thus, among horses whose odds decline by 10\% during the final five minutes and whose final odds are $R=20$, the predicted average gross return is approximately 8.4 percentage points higher when the preceding 10\% movement is an odds increase (reversal) rather than an odds decline (continuation).

Next, consider horses whose odds increase by $d$ during the final five minutes, so that $\Delta_{i,L}^{+}=d$ and $\Delta_{i,L}^{-}=0$. Under reversal, odds decline by $d$ during $[-10,-5]$, so that $\Delta_{i,E}^{-}=-d$ and $\Delta_{i,E}^{+}=0$. Under continuation, odds increase by $d$ over the same interval, so that $\Delta_{i,E}^{+}=d$ and $\Delta_{i,E}^{-}=0$. The corresponding expected-return difference is
\[
\begin{aligned}
&
\mathbb{E}\!\left[
\mathbf{1}_{\{win_i=1\}}R_i^{\ast}
\mid
R_i^{\ast}=R,\,
\Delta_{i,L}^{+}=d,\,
\Delta_{i,E}^{-}=-d
\right]
\\
&\quad-
\mathbb{E}\!\left[
\mathbf{1}_{\{win_i=1\}}R_i^{\ast}
\mid
R_i^{\ast}=R,\,
\Delta_{i,L}^{+}=d,\,
\Delta_{i,E}^{+}=d
\right]
\\
&=
-d(\delta_E^{+}+\delta_E^{-})
-
Rd(\gamma_E^{+}+\gamma_E^{-})
-
d^2\kappa_{-+}
-
Rd^2\lambda_{-+}.
\end{aligned}
\]

Evaluating this expression at $R=20$ and $d=0.1$ using the point estimates from the rich-fixed-effects specification (column~(3) of Table~\ref{tb:estimation_results_odds_extended_symmetry}) gives
\[
\begin{aligned}
&-0.1(0.03949+0.64506)
-20(0.1)(-0.00072+0.00027)\\
&\qquad
-(0.1)^2(-1.53154)
-20(0.1)^2(0.00497)
=-0.0532.
\end{aligned}
\]
Thus, among horses whose odds increase by 10\% during the final five minutes and whose final odds are $R=20$, the predicted average gross return is approximately 5.3 percentage points lower when the preceding 10\% movement is an odds decline (reversal) rather than an odds increase (continuation).

\clearpage
\section{Additional Testable Implications}
 \renewcommand\theequation{\thesection.\arabic{equation}}
 \renewcommand\thefigure{\thesection.\arabic{figure}} 
  \renewcommand\thetable{\thesection.\arabic{table}} 
  \setcounter{equation}{0}
 \setcounter{figure}{0}
  \setcounter{table}{0}
\subsection{Racetrack Conditions}\label{sec:add_test_racetrack}

We examine whether the strength of the final-odds gradient and the path dependence in returns vary with racetrack conditions. To do so, we use the racetrack-condition information reported in the JRA data and group the reported categories into three indicators: normal conditions, slightly bad conditions, and bad conditions. Normal conditions are used as the reference group. We then estimate an interaction specification in which these condition indicators are interacted with final odds, final-five-minute odds changes, and their interaction.

Table~\ref{tb:estimation_results_odds_interact_racetrack_condition} reports the results. The coefficient on final odds remains negative under normal racetrack conditions, reproducing the conventional FLB. The interaction terms between final odds and adverse racetrack conditions are positive, especially for bad conditions, indicating that the negative final-odds gradient is attenuated when racetrack conditions deteriorate.\footnote{Under slightly bad conditions, the linear final-odds gradient, evaluated at zero odds change, remains negative but smaller in magnitude; under bad conditions, its point estimate becomes positive, indicating a further attenuation of the FLB.} The coefficients on final-five-minute odds changes remain negative. The interactions with adverse conditions are positive and sizable, although statistically insignificant in many specifications. Thus, the estimates are suggestive of attenuation in the association between late odds declines and higher realized returns under poorer track conditions. Overall, these patterns are consistent with the implication that noisier racing environments attenuate the strength of the FLB by weakening the informativeness of private signals.

\begin{table}[t!]
  \footnotesize
  \caption{Return Predictability and Racetrack Condition}
  \label{tb:estimation_results_odds_interact_racetrack_condition}
  \begin{center}
      \adjustbox{max width=0.9\textwidth}{
\begin{tabular}[t]{lccc}
\toprule
  & (1) No FE & (2) Race FE & (3) Rich FE\\
\midrule
$R_{i}^{*}$ & -0.00155 & -0.00135 & -0.00134\\
 & (0.00014) & (0.00019) & (0.00019)\\
$R_{i}^{*} \times D_r^{-}$ & 0.00065 & 0.00070 & 0.00060\\
 & (0.00033) & (0.00046) & (0.00043)\\
$R_{i}^{*} \times D_r^{--}$ & 0.00328 & 0.00346 & 0.00328\\
 & (0.00046) & (0.00167) & (0.00150)\\
$\frac{\Delta R_{i,[-5, 0]}}{R_{i,-5}}$ & -0.39543 & -0.39064 & -0.37837\\
 & (0.04504) & (0.04611) & (0.04580)\\
$\frac{\Delta R_{i,[-5, 0]}}{R_{i,-5}} \times D_r^{-}$ & 0.14330 & 0.11579 & 0.14469\\
 & (0.10514) & (0.10360) & (0.10322)\\
$\frac{\Delta R_{i,[-5, 0]}}{R_{i,-5}} \times D_r^{--}$ & 0.10338 & 0.09432 & 0.10603\\
 & (0.14236) & (0.18231) & (0.16718)\\
$R_{i}^{*} \times \frac{\Delta R_{i,[-5, 0]}}{R_{i,-5}}$ & 0.00089 & 0.00046 & 0.00074\\
 & (0.00030) & (0.00040) & (0.00037)\\
$R_{i}^{*} \times \frac{\Delta R_{i,[-5, 0]}}{R_{i,-5}} \times D_r^{-}$ & -0.00113 & -0.00117 & -0.00107\\
 & (0.00070) & (0.00083) & (0.00076)\\
$R_{i}^{*} \times \frac{\Delta R_{i,[-5, 0]}}{R_{i,-5}} \times D_r^{--}$ & -0.00484 & -0.00532 & -0.00483\\
 & (0.00100) & (0.00230) & (0.00200)\\
$D_r^{-}$ & -0.01192 &  & -0.01312\\
 & (0.02240) &  & (0.01209)\\
$D_r^{--}$ & -0.07258 &  & -0.07661\\
 & (0.03028) &  & (0.03969)\\
\midrule
Num.Obs. & 894127 & 894127 & 894127\\
R2 & 0.001 & 0.054 & 0.002\\
R2 Adj. & 0.001 & -0.018 & 0.001\\
FE: Race &  & X & \\
FE: Race-Year &  &  & X\\
FE: Race-Location &  &  & X\\
FE: Race-Grade &  &  & X\\
FE: Jockey &  &  & X\\
Horse Characteristics &  &  & X\\
\bottomrule
\end{tabular}

      }
  \end{center}

     \begin{tablenotes}[flushleft]
\footnotesize
\item \textit{Notes:} \(D_r^-\) and \(D_r^{--}\) are dummy variables for slightly bad and bad racetrack conditions in race \(r\), respectively, with normal conditions as the omitted category. The dependent variable is \(R_i^\ast \times \mathbf{1}_{\{win_i=1\}}\). Standard errors are clustered at the race level in Columns~(2)--(3).  In Column~(2), the racetrack-condition main effects are absorbed by race fixed effects.
\end{tablenotes}  
\end{table}

\subsection{Participation Margin and Late Betting Share}
\label{sec:add_test_participant}

We examine whether the association between odds changes and subsequent returns varies with the amount of late betting activity. The motivation comes from the participation margin in the two-period model discussed in Section~\ref{Sec:Discuss}. When the private value of betting is finite, bettors with weak signals may abstain, so the amount of observed late betting may reflect not only the number of informed bettors but also selection into participation. To explore this channel, we use the share of bets placed in the final five minutes as a proxy for late participation. Specifically, we define \(D_r^{\mathrm{HighLate}}\) as a dummy variable equal to one for races in which the final-five-minute betting share exceeds the sample median, 0.468.

Table~\ref{tab:letter_sq50_oddschange} reports regressions in which the coefficients on final odds, odds changes, and their interaction are allowed to differ by \(D_r^{\mathrm{HighLate}}\). Across specifications, the coefficient on \(R_i^\ast\times D_r^{\mathrm{HighLate}}\) is positive, indicating that the FLB is weaker when the final-five-minute betting share is high. The coefficient on the interaction between odds changes and the high-share dummy is also positive, suggesting that the predictive content of odds declines is weaker in high late-betting-share races.

These patterns are consistent with the participation-selection interpretation discussed in the main text. A high final-five-minute betting share need not imply stronger information aggregation if it also includes more weak-signal bettors who would otherwise abstain. Conversely, when late participation is low, observed late bettors may be more strongly selected on signal strength, making late odds movements more informative about subsequent returns.

\begin{table}[t!]
\caption{Return Predictability and Last-Five-Minute Betting Share}
\label{tab:letter_sq50_oddschange}
  \begin{center}
\adjustbox{max width=\textwidth}{%

\begin{tabular}[t]{lccc}
\toprule
  & No FE & Race FE & Rich FE\\
\midrule
$R_{i}^{*}$ & -0.00127 & -0.00115 & -0.00110\\
 & (0.00018) & (0.00026) & (0.00026)\\
$R_{i}^{*} \times D_r^{\mathrm{HighLate}}$ & 0.00066 & 0.00086 & 0.00069\\
 & (0.00027) & (0.00047) & (0.00045)\\
$\frac{\Delta R_{i,[-5, 0]}}{R_{i,-5}}$ & -0.53717 & -0.53766 & -0.50961\\
 & (0.07573) & (0.07501) & (0.07154)\\
$\frac{\Delta R_{i,[-5, 0]}}{R_{i,-5}} \times D_r^{\mathrm{HighLate}}$ & 0.27828 & 0.28580 & 0.26296\\
 & (0.08944) & (0.08984) & (0.08624)\\
$R_{i}^{*} \times \frac{\Delta R_{i,[-5, 0]}}{R_{i,-5}}$ & 0.00024 & -0.00013 & 0.00016\\
 & (0.00058) & (0.00084) & (0.00080)\\
$R_{i}^{*} \times \frac{\Delta R_{i,[-5, 0]}}{R_{i,-5}} \times D_r^{\mathrm{HighLate}}$ & -0.00113 & -0.00134 & -0.00113\\
 & (0.00068) & (0.00102) & (0.00096)\\
$D_r^{\mathrm{HighLate}}$ & -0.01122 &  & -0.01312\\
 & (0.01660) &  & (0.01261)\\
\midrule
Num.Obs. & 833700 & 833700 & 833700\\
R2 & 0.001 & 0.054 & 0.002\\
R2 Adj. & 0.001 & -0.018 & 0.001\\
FE: Race &  & X & \\
FE: Race-Year &  &  & X\\
FE: Race-Location &  &  & X\\
FE: Race-Grade &  &  & X\\
FE: Jockey &  &  & X\\
Horse Characteristics &  &  & X\\
\bottomrule
\end{tabular}
}
  \end{center}
     \begin{tablenotes}[flushleft]
\footnotesize
\item \textit{Notes:}  The dummy variable \(D_r^{\mathrm{HighLate}}\) is equal to one when race $r$'s share of bets placed in the final five minutes exceeds the sample median ($0.468$). The dependent variable is $R_i^\ast \times \mathbf{1}_{\{win_i=1\}}$. Standard errors are clustered at the race level in Columns~(2)--(3). 
\end{tablenotes}
\end{table}

\clearpage
\subsection{Quinella Odds Analysis}\label{App:Quinella}
Table~\ref{tb:summary_statistics_horse_race_quinella} reports summary statistics 
for quinella odds from 2019 to 2023, with the shorter sample period chosen for 
computational feasibility (in contrast, the win odds data in 
Table~\ref{tb:summary_statistics_horse_race} cover 2004--2023). Compared with the win odds, two differences stand out. First, Panel~(a) shows that the total amount wagered in quinella bets is larger on average (1.16 million JPY per race) than in win bets (0.40 million JPY). Yet, because the quinella market involves far more betting objects---around 92 horse pairs per race compared with about 14 horses---the amount wagered per pair is much smaller than the amount wagered per horse in win betting. Panel~(b) shows that quinella odds are substantially larger, averaging over 500 for horse-pair-race observations, whereas final win odds average only about 67.

\begin{table}[h!]
  \begin{center}
  \caption{Summary Statistics of Quinella Odds Data}
  \label{tb:summary_statistics_horse_race_quinella}
  \subfloat[Race level]{
  
\begin{tabular}[t]{lrrrrr}
\toprule
  & N & Mean & SD & Min. & Max.\\
\midrule
Num Horse Pairs & 14289 & 92.06 & 34.72 & 10.00 & 153.00\\
Num. of Wagers (Mil) & 14289 & 1.16 & 2.31 & 0.14 & 67.37\\
\bottomrule
\end{tabular}

  }\\
  \subfloat[Horse-pair-race level]{
  
\begin{tabular}[t]{lrrrrr}
\toprule
  & N & Mean & SD & Min. & Max.\\
\midrule
Final Odds & 1315425 & 513.33 & 791.98 & 1.10 & 74093.80\\
\bottomrule
\end{tabular}

  }
  \end{center}
  \footnotesize
  Note: The quinella odds data cover all centrally administered races in Japan from 2019 to 2023. 
\end{table}

\begin{figure}[t!]
          \caption{Quinella Odds versus Realized Returns}
            \centering
    \includegraphics[width = .8\textwidth]{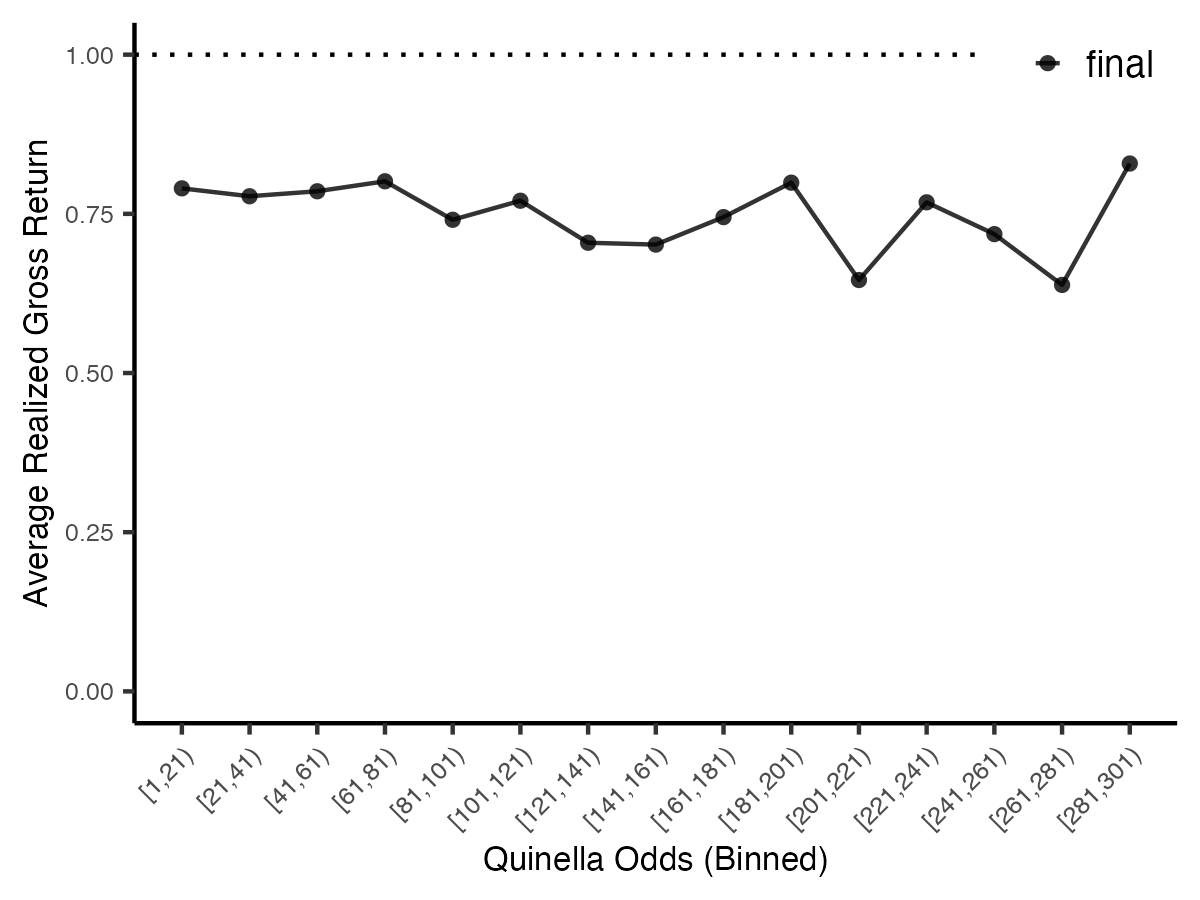}
    \label{fg:plot_expected_return_intermediate_quinella_odds_min_before}
    
         \begin{tablenotes}[flushleft]
\footnotesize
\item \textit{Notes:} The average realized gross return for each group \( g \in \mathcal{G} \) on the vertical axis is given by $\frac{1}{| I_g |} \sum_{i \in I_g} \mathbf{1}_{\{win_i = 1\}} R_i^{\ast}$, where \( I_g \) denotes the set of horse pairs in group \( g \), and \( R_i^{\ast} \) represents the final odds for horse pair \( i \). 
    \end{tablenotes}
\end{figure}

Figure \ref{fg:plot_expected_return_intermediate_quinella_odds_min_before} presents an analogous relationship between quinella odds and realized returns. Unlike the monotonic pattern observed in win odds in Figure~\ref{fg:plot_race_odds_return}, the relationship for quinella bets appears less systematic, suggesting that the classic FLB is more muted in exotic bets such as quinella.

\begin{figure}[t!]
  \caption{Expected Return versus Interim Quinella Odds Changes}
  \label{fg:plot_expected_return_intermediate_quinella_odds_min_before_5_to_0} 
  \begin{center}
  \subfloat[-5 to 0 (final)]{\includegraphics[width = 0.36\textwidth]{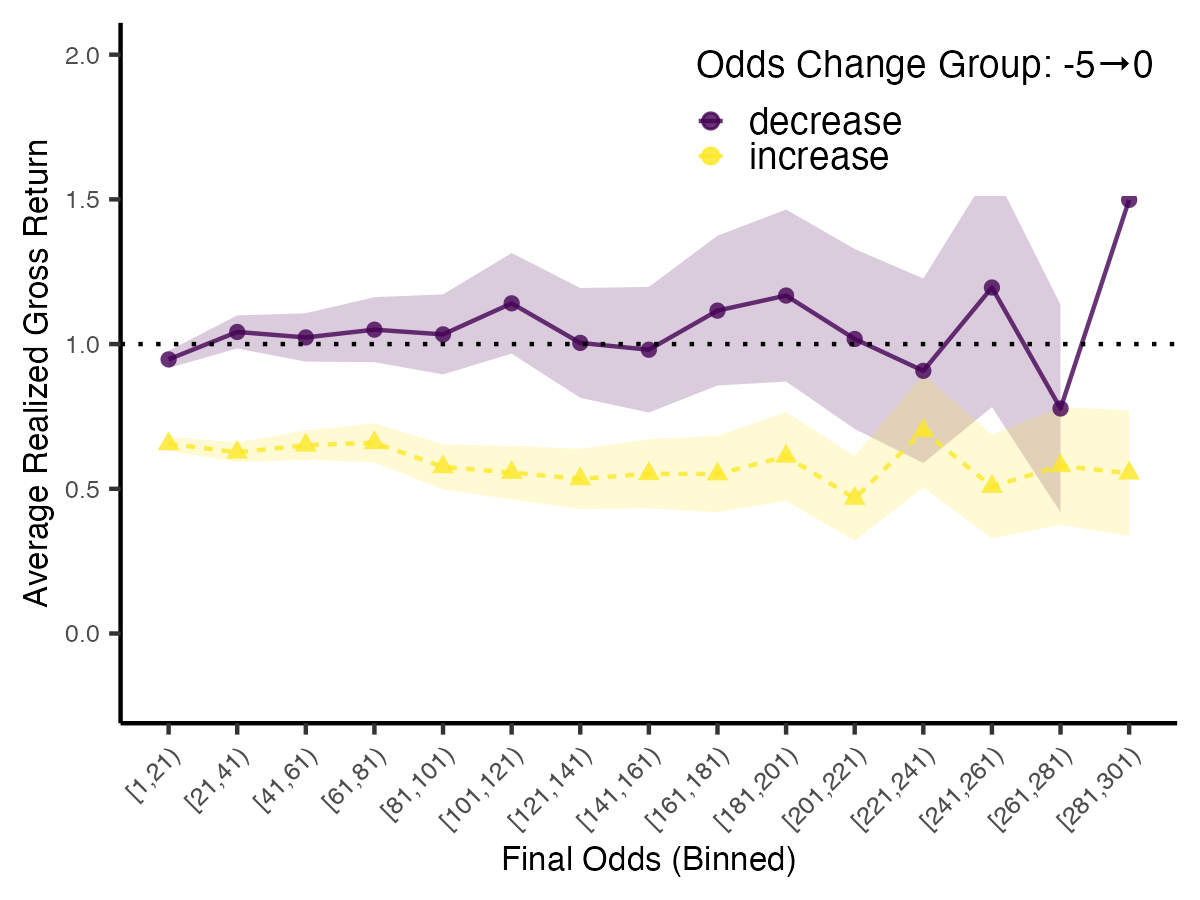}}
  \qquad\qquad
  \subfloat[-10 to -5]{\includegraphics[width = 0.36\textwidth]{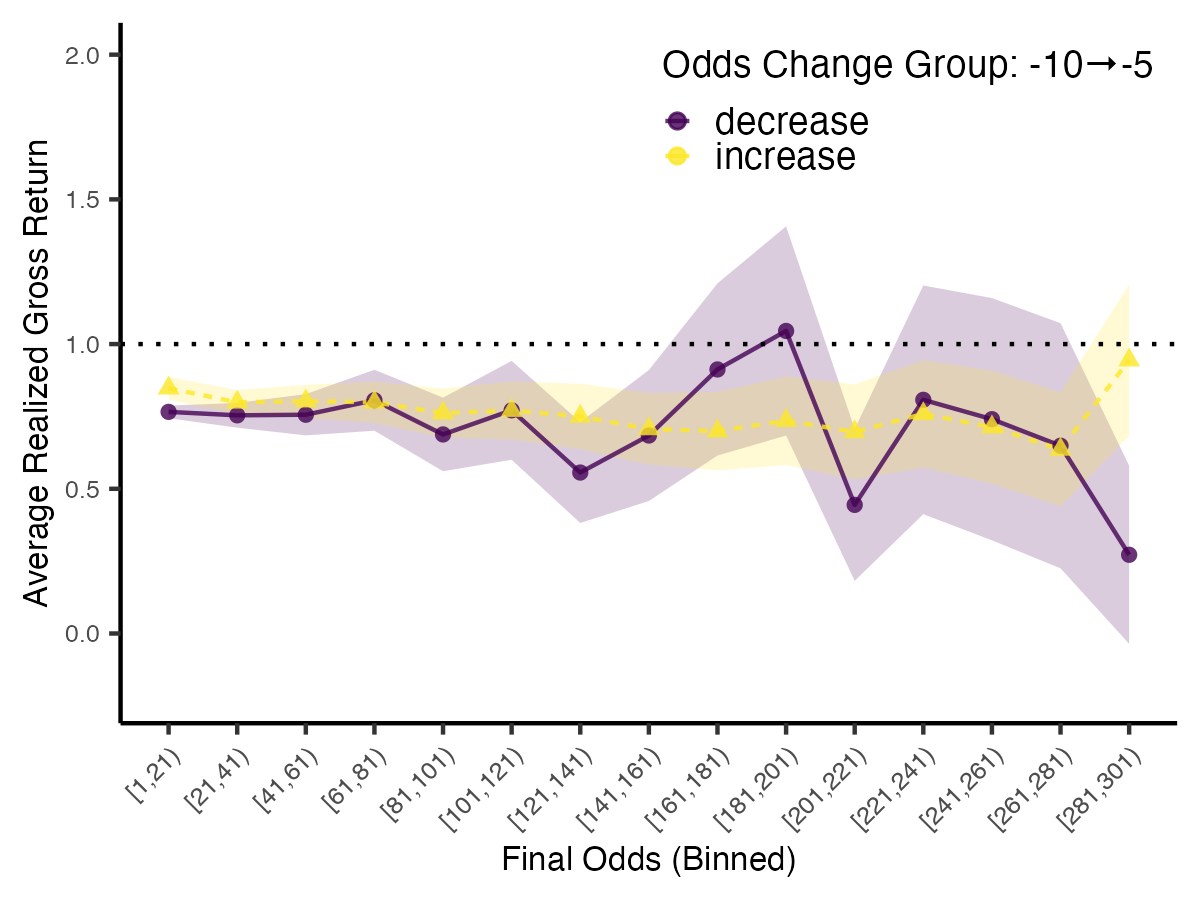}}\\
  \subfloat[-15 to -10]{\includegraphics[width = 0.36\textwidth]{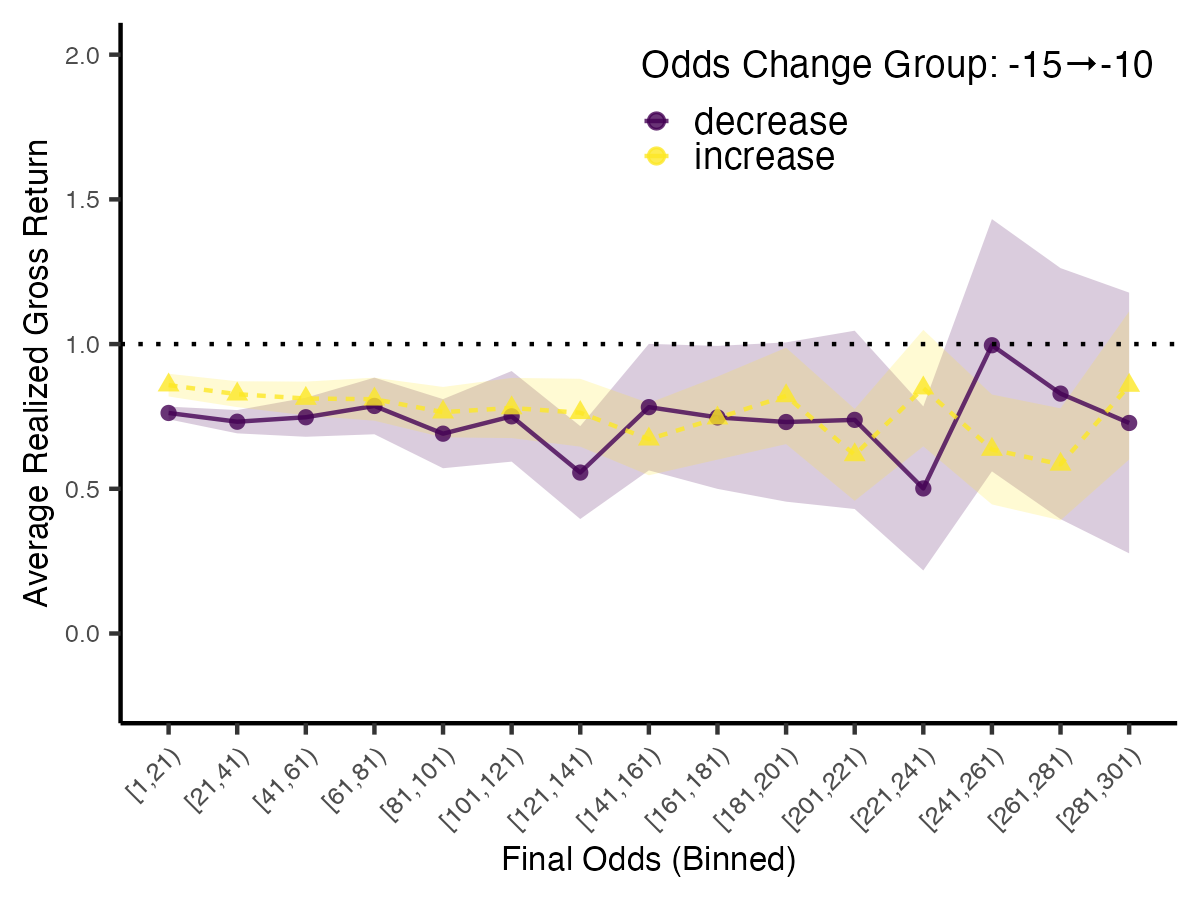}}
  \qquad\qquad
  \subfloat[-20 to -15]{\includegraphics[width = 0.36\textwidth]{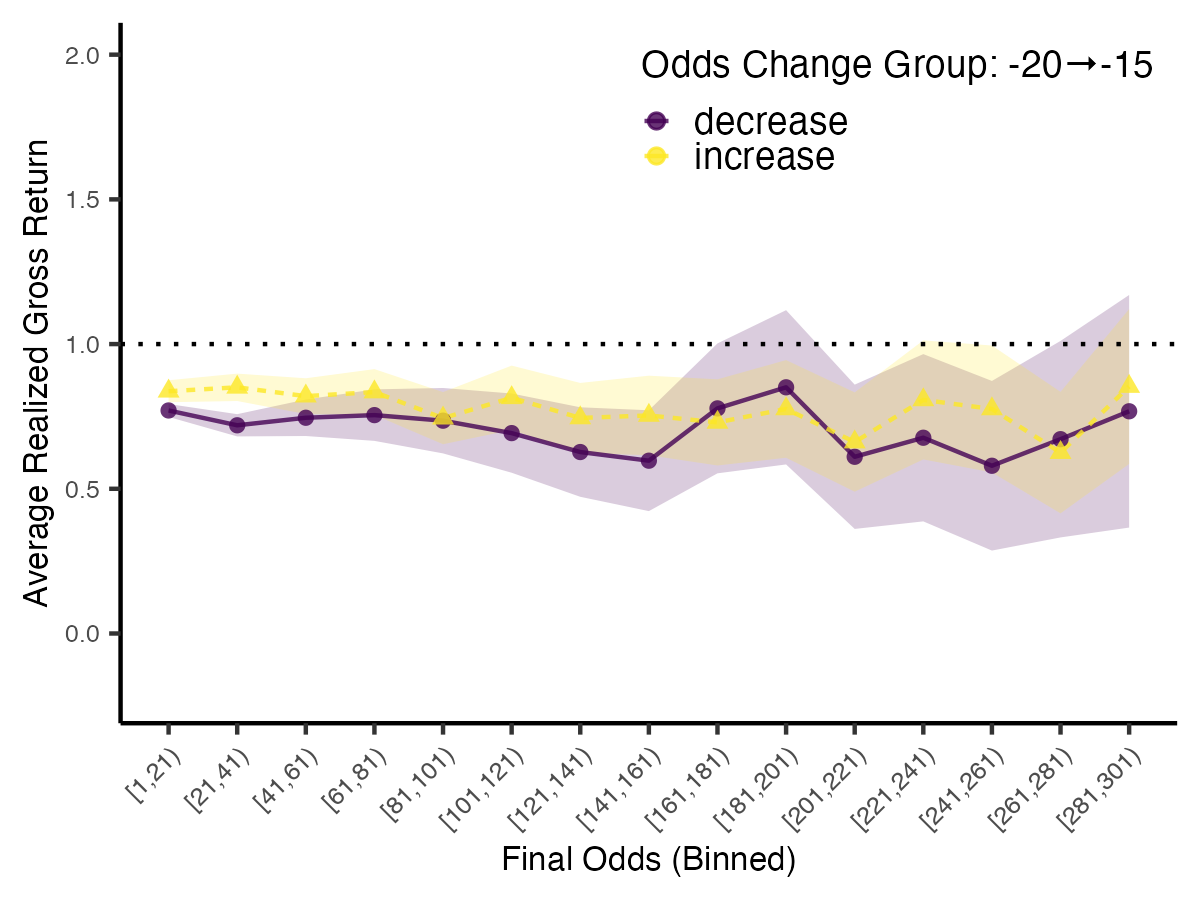}}\\
  \subfloat[-25 to -20]{\includegraphics[width = 0.36\textwidth]{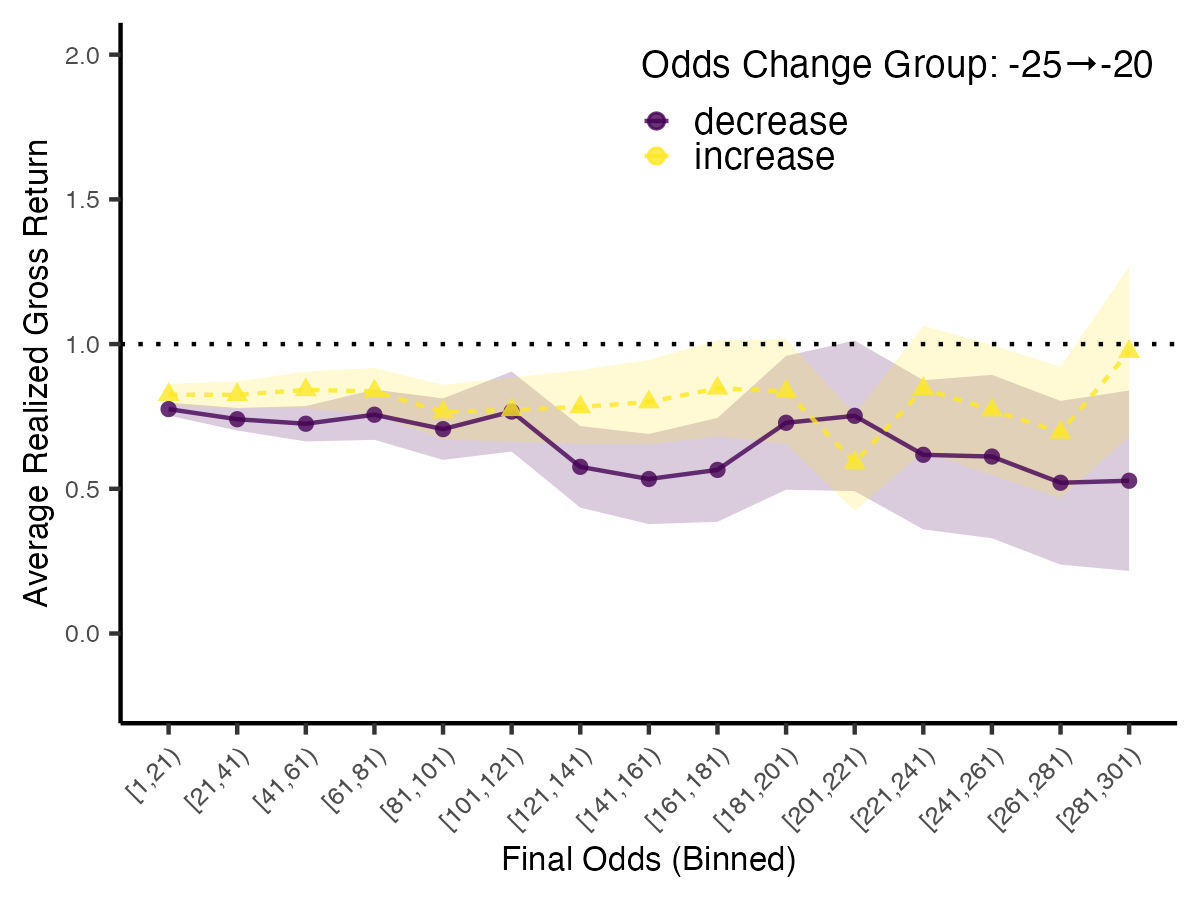}}
  \qquad\qquad
  \subfloat[-30 to -25]{\includegraphics[width = 0.36\textwidth]{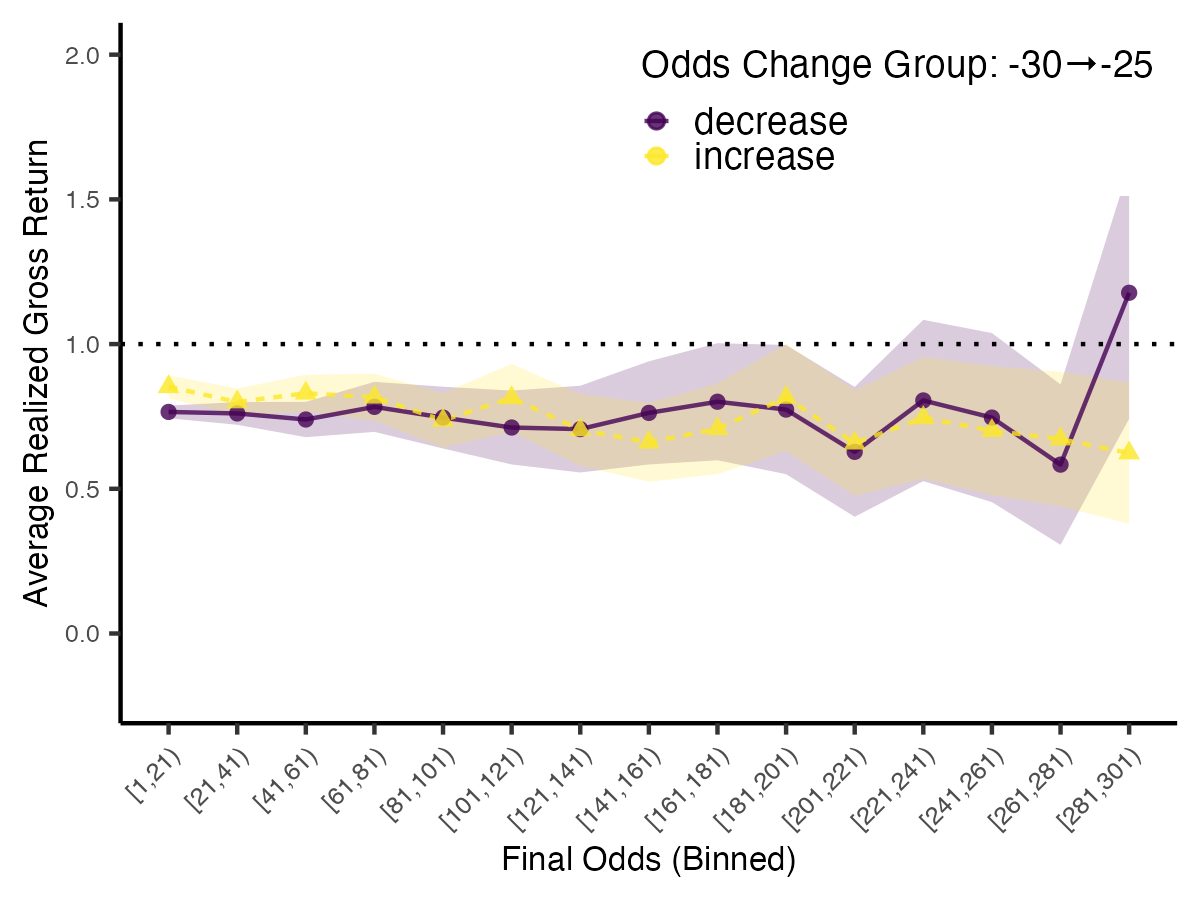}}
  
  \end{center}
  \footnotesize
  \textit{Notes:} The horizontal axis represents final quinella odds, grouped into bins of width 20. Panels (a)--(f) compare average realized returns for horse pairs whose odds increased versus decreased during specified five-minute intervals prior to post time. Panel (a) covers the final interval (0 to 5 minutes before post time), while Panels (b)--(f) examine earlier intervals: 5 to 10, 10 to 15, 15 to 20, 20 to 25, 25 to 30 minutes before post time, respectively. Within each panel, the two lines plot average realized returns for horse pairs with increasing and decreasing odds during the corresponding interval. Shaded areas denote 95\% confidence intervals. Following \citet{jullien2000estimating}, bins are excluded if either subgroup contains fewer than 1,000 observations.
\end{figure}

Figure \ref{fg:plot_expected_return_intermediate_quinella_odds_min_before_5_to_0} 
examines whether changes in interim quinella odds affect realized returns, mirroring the analysis for win bets in Figure~\ref{fg:plot_expected_return_intermediate_odds_min_before_5_to_0}. 
Across all panels, we compare average returns conditional on final odds for bets whose interim odds either increased or decreased over successive five-minute intervals before post time. 

As in the win bet case, we observe a divergence between the ``increase'' and ``decrease'' groups, with odds declines associated with higher realized returns and odds increases with lower returns. This pattern becomes clearly visible only in the final five minutes (Panel~(a)), consistent with the pattern observed in win odds. Overall, while the classic FLB appears more muted in quinella bets, such bets nevertheless exhibit a similar form of late-stage predictability to that in win markets.

\begin{table}[t!]
  \footnotesize
  \caption{Estimation Results: Quinella Odds}
  \label{tb:estimation_results_quinella_odds}
  \begin{center}
      
\begin{tabular}[t]{lcccccc}
\toprule
  & (1) & (2) & (3) & (4) & (5) & (6)\\
\midrule
Constant & 0.69829 & 0.77976 & 0.80470 & 0.80601 & 0.80031 & 0.76995\\
 & (0.01584) & (0.01711) & (0.01778) & (0.01802) & (0.01820) & (0.01893)\\
$R_{i}^{*}$ & -0.00011 & -0.00012 & -0.00013 & -0.00012 & -0.00012 & -0.00013\\
 & (0.00001) & (0.00002) & (0.00002) & (0.00002) & (0.00002) & (0.00002)\\
$\frac{\Delta R_{i,[-5, 0]}}{R_{i,-5}}$ &  & -1.40233 &  &  &  & -1.42629\\
 &  & (0.08494) &  &  &  & (0.08852)\\
$\frac{\Delta R_{i,[-10, 0]}}{R_{i,-10}}$ &  &  & -0.97162 &  &  & \\
 &  &  & (0.06676) &  &  & \\
$\frac{\Delta R_{i,[-15, 0]}}{R_{i,-15}}$ &  &  &  & -0.79996 &  & \\
 &  &  &  & (0.05799) &  & \\
$\frac{\Delta R_{i,[-20, 0]}}{R_{i,-20}}$ &  &  &  &  & -0.59334 & \\
 &  &  &  &  & (0.04879) & \\
$\frac{\Delta R_{i,[-10, -5]}}{R_{i,-10}}$ &  &  &  &  &  & 0.32016\\
 &  &  &  &  &  & (0.22851)\\
$\frac{\Delta R_{i,[-15, -10]}}{R_{i,-15}}$ &  &  &  &  &  & -0.35772\\
 &  &  &  &  &  & (0.32021)\\
$\frac{\Delta R_{i,[-20, -15]}}{R_{i,-20}}$ &  &  &  &  &  & 0.35256\\
 &  &  &  &  &  & (0.24231)\\
$R_{i}^{*} \times \frac{\Delta R_{i,[-5, 0]}}{R_{i,-5}}$ &  & 0.00023 &  &  &  & 0.00022\\
 &  & (0.00005) &  &  &  & (0.00005)\\
$R_{i}^{*} \times \frac{\Delta R_{i,[-10, 0]}}{R_{i,-10}}$ &  &  & 0.00018 &  &  & \\
 &  &  & (0.00003) &  &  & \\
$R_{i}^{*} \times \frac{\Delta R_{i,[-15, 0]}}{R_{i,-15}}$ &  &  &  & 0.00013 &  & \\
 &  &  &  & (0.00002) &  & \\
$R_{i}^{*} \times \frac{\Delta R_{i,[-20, 0]}}{R_{i,-20}}$ &  &  &  &  & 0.00010 & \\
 &  &  &  &  & (0.00002) & \\
$R_{i}^{*} \times \frac{\Delta R_{i,[-10, -5]}}{R_{i,-10}}$ &  &  &  &  &  & 0.00012\\
 &  &  &  &  &  & (0.00008)\\
$R_{i}^{*} \times \frac{\Delta R_{i,[-15, -10]}}{R_{i,-15}}$ &  &  &  &  &  & -0.00008\\
 &  &  &  &  &  & (0.00011)\\
$R_{i}^{*} \times \frac{\Delta R_{i,[-20, -15]}}{R_{i,-20}}$ &  &  &  &  &  & -0.00005\\
 &  &  &  &  &  & (0.00009)\\
\midrule
Num.Obs. & 1315425 & 1315425 & 1315425 & 1315425 & 1315425 & 1315425\\
R2 & 0.000 & 0.000 & 0.000 & 0.000 & 0.000 & 0.000\\
R2 Adj. & 0.000 & 0.000 & 0.000 & 0.000 & 0.000 & 0.000\\
\bottomrule
\end{tabular}

  \end{center}
  
     \begin{tablenotes}[flushleft]
\footnotesize
\item \textit{Notes:} This table reports OLS estimates of the coefficients from the regression:  
\[
 \mathbf{1}_{\{win_{i}=1\}}R_{i}^{\ast} = \alpha + \beta R_{i}^{\ast} + \delta \,  \text{OddsChange}_{i} + \gamma \,  R_{i}^{\ast} \times \text{OddsChange}_{i} + \varepsilon_{i}.
\]  
The variable \( \Delta R_{i,[-\tau,0]}/R_{i,-\tau} \equiv (R_{i}^{\ast} - R_{i,-\tau}) / R_{i,-\tau} \) denotes the rate of change in odds over the final \( \tau \) minutes before post time.  
    \end{tablenotes}
\end{table}

\clearpage 
\section{Two-Period Betting Model}
 \renewcommand\theequation{\thesection.\arabic{equation}}
 \renewcommand\thefigure{\thesection.\arabic{figure}} 
  \renewcommand\thetable{\thesection.\arabic{table}} 
  \setcounter{equation}{0}
 \setcounter{figure}{0}
  \setcounter{table}{0}

\subsection{Biased First-Period Betting Shares}\label{App:BiasedFirstPeriodShares}
\begin{figure}[h!]
          \caption{Expected Returns When First-period Bettors Are Biased}
            \centering
    \includegraphics[width = .8\textwidth]{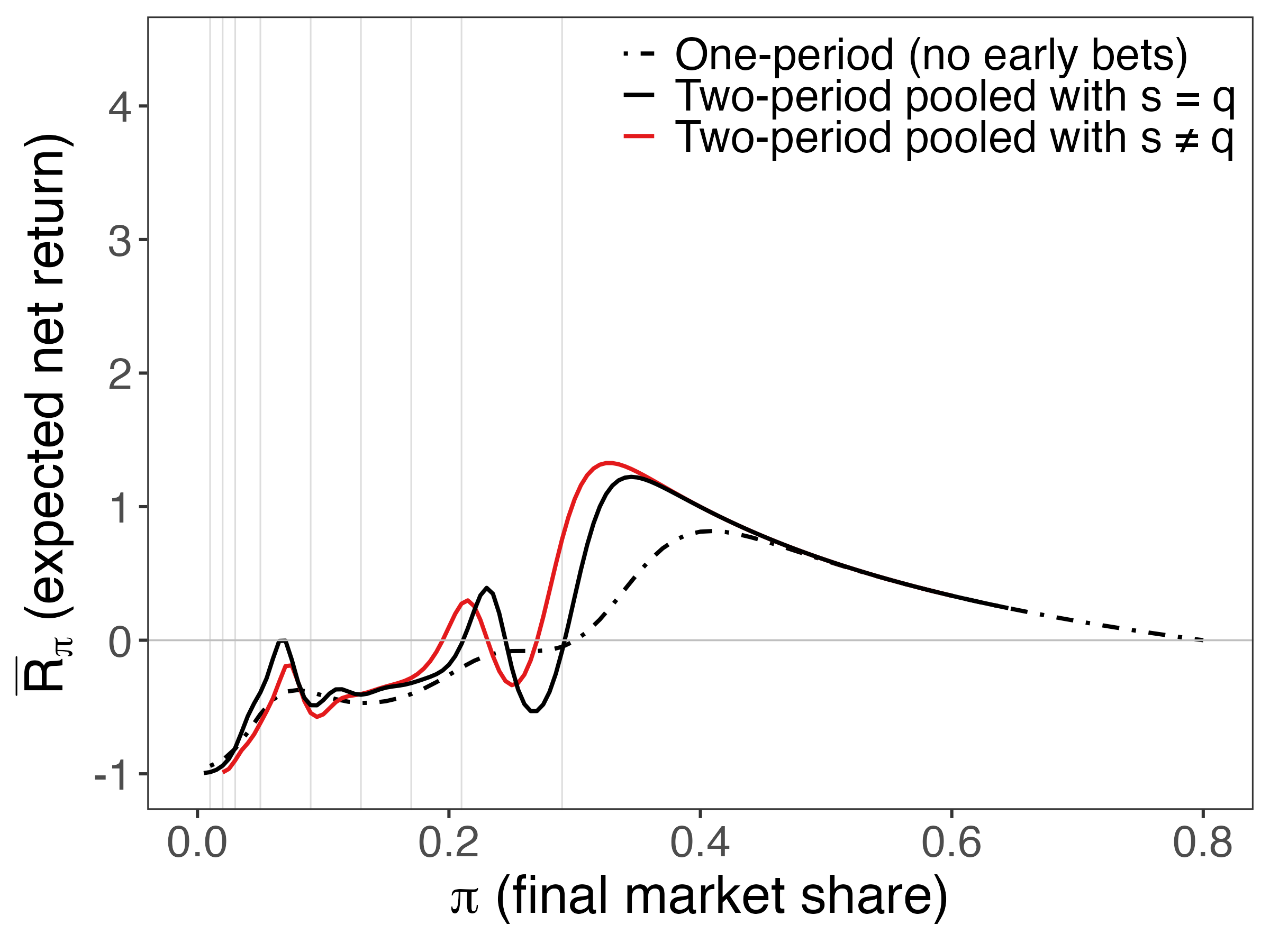}
    \label{fg:return_curve_two_period_vs_one_period_flb}
    
         \begin{tablenotes}[flushleft]
\footnotesize
\item \textit{Notes:} The figure plots the expected (net) return $\overline{R}_{\pi}$ against the final market share $\pi$ for a nine-horse race with objective winning probabilities 
$q=(0.29,0.21,0.17,0.13,0.09,0.05,0.03,0.02,0.01)$ and a takeout rate of 20\%. 
It compares a one-period benchmark with 100 informed bettors who bet simultaneously to two different settings in a two-period model with 100 pre-existing first-period bets and 100 informed second-period bettors. The black solid line sets the first-period market shares equal to the objective probabilities, $s_i=q_i$. The red solid line allows the first-period shares to exhibit an FLB by setting $s_i=q_i^{\alpha}/\sum_{j=1}^{9}q_j^{\alpha}$ with $\alpha = 0.5$. Vertical gray lines mark the objective winning probabilities $q_i$.
    \end{tablenotes}
\end{figure}

The baseline simulation sets the first-period market shares equal to the objective winning probabilities, $s=q$, to isolate the role of late information aggregation. As a robustness exercise, we instead allow the first-period shares to exhibit a favorite--longshot bias, so that $s_i=q_i^{\alpha}/\sum_{j=1}^{K}q_j^{\alpha}$ with $\alpha =0.5$, while holding the remaining parameters fixed. As shown in Figure~\ref{fg:return_curve_two_period_vs_one_period_flb}, biased early betting changes the expected-return schedule quantitatively, but the qualitative contrast between the two-period information model and the one-period benchmark remains. Thus, the simulation results do not depend on the assumption that first-period market shares coincide with objective winning probabilities.

\subsection{Risk-Preference Estimation in Simulated Data}
\label{App:RiskPreferenceSimulation}
 \renewcommand\theequation{\thesection.\arabic{equation}}
 \renewcommand\thefigure{\thesection.\arabic{figure}} 
  \renewcommand\thetable{\thesection.\arabic{table}} 
  \setcounter{equation}{0}
 \setcounter{figure}{0}
  \setcounter{table}{0}

This appendix examines whether an information-based betting model can generate patterns that are captured by standard structural models of risk preferences. The motivation is the misspecification concern discussed in Section~\ref{Sec:Discuss}: if final odds reflect not only preferences or subjective probability distortions but also private-information aggregation and the timing of informed betting, structural estimates based only on final odds and realized outcomes may partly absorb informational variation into preference parameters.

We conduct a simulation exercise based on the two-period betting model. Each simulated race has nine horses. For each race, we first draw objective winning probabilities \(q\) from a Dirichlet distribution centered around
\(\bar{q}=(0.29,0.21,0.17,0.13,0.09,0.05,0.03,0.02,0.01)\). Given \(q\), we solve the two-period betting model and simulate final market shares, final odds, and the realized winner. Repeating this procedure generates a synthetic dataset of \(10{,}000\) race outcomes in which the data-generating process is driven by information aggregation rather than heterogeneity in risk preferences. 

\begin{figure}[t!]
  \begin{center}
  \caption{Analytical and Simulated Return Patterns}
  \label{fg:plot_simulation_curve} 
  \includegraphics[width = 0.8\textwidth]{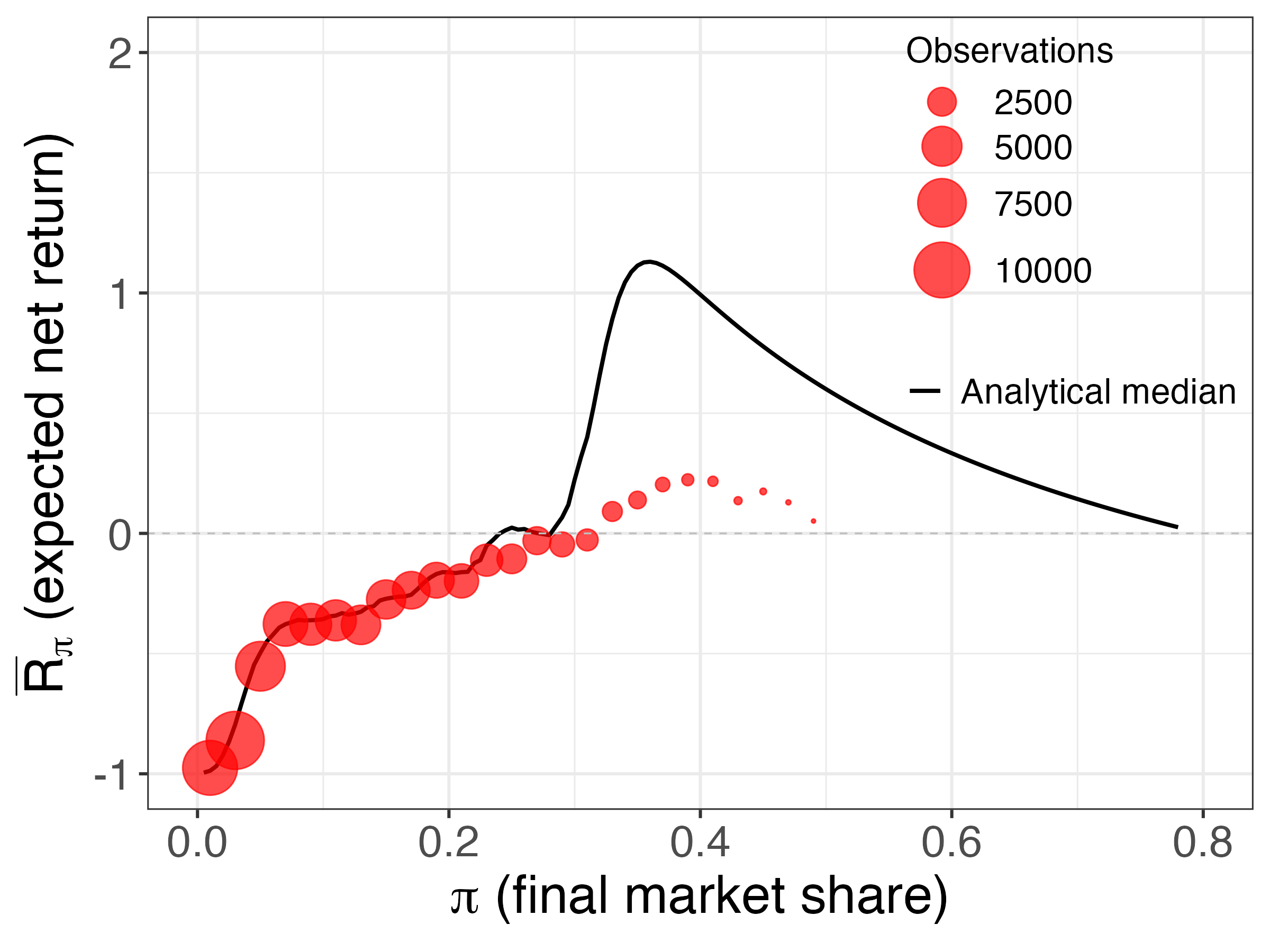}
  \end{center}
         \begin{tablenotes}[flushleft]
        \footnotesize
    \item \textit{Notes:} For each of 10,000 draws, the objective winning probability vector \(q\) is sampled from \(\mathrm{Dirichlet}(\kappa \bar q)\), with \(\kappa=50\) and \(\bar q=(0.29,0.21,0.17,0.13,0.09,0.05,0.03,0.02,0.01)\). The first-period market shares are set equal to \(s=q\), and the Bayesian Nash equilibrium of the two-period betting model is solved numerically for each draw. The solid line plots the median analytical expected return \(\overline R_{\pi}\) across draws, evaluated at feasible final market shares. Red circles plot average realized returns from Monte Carlo simulations, pooling horse-level outcomes and binning them by final market share with bin width 0.02. Circle size is proportional to the number of observations in each bin. Parameters are \(\theta=100\), \(B_0=100\), \(M=100\), and \(\rho=0.20\).
    \end{tablenotes}
\end{figure}

As a diagnostic check, Figure~\ref{fg:plot_simulation_curve} compares the analytical expected-return curve with Monte Carlo outcomes generated from the same two-period model. The simulated bin averages reproduce the main qualitative pattern of the analytical curve, indicating that the simulation procedure used for the preference-estimation exercise is consistent with the model-implied return patterns.

We then apply the estimation approach of \citet{chiappori2019aggregate} to the simulated data. Specifically, we estimate the same class of structural specifications considered in their analysis, including expected utility and rank-dependent expected utility models, allowing for homogeneous and heterogeneous bettor types. The purpose of the exercise is not to estimate preferences from actual betting behavior, but to ask whether data generated by an information-based mechanism can be well described by preference-based models.

Table~\ref{tab:mc_all_bic_model_selection} reports the model comparison results. The homogeneous rank-dependent expected utility specification provides the best fit to the simulated data. This finding is noteworthy because \citet{chiappori2019aggregate} also find that the homogeneous rank-dependent expected utility model best fits their field data. In our simulation, however, the data are generated without imposing rank-dependent preferences or probability weighting as primitives. The result therefore suggests that dynamic information aggregation can generate odds--return patterns that resemble those typically attributed to preference distortions.

This exercise should be interpreted as suggestive rather than as a direct critique of structural preference estimation. Nevertheless, it highlights that an omitted information channel may affect the interpretation of estimated preference parameters. In particular, when odds incorporate private information through dynamic betting, structural models that do not explicitly account for this channel may attribute part of the information-driven odds--return relationship to risk preferences, probability weighting, or subjective beliefs.

\begin{table}[t!]
  \footnotesize
  \caption{Model Comparison of Risk-Preference Estimation in Simulated Data}
  \label{tab:mc_all_bic_model_selection}
  \begin{center}
  \begin{tabular}[t]{lcccc}
\toprule
Model family & $V(R,p,\vartheta)$ & BIC & No. parameters & $2\log L$\\
\midrule
Risk-neutral & $p(1+R)$ & -32,000.2 & 0 & -32,000.2\\
EU homogeneous & $p\,u(R)$ & -31,130.1 & 3 & -31,102.5\\
EU heterogeneous & $p\,u(R,\vartheta)$ & -31,096.0 & 4 & -31,059.1\\
Yaari homogeneous & $G(p)(1+R)$ & -31,187.4 & 4 & -31,150.6\\
RDEU homogeneous & $G(p)\,u(R)$ & -31,038.2 & 2 & -31,019.8\\
NEU homogeneous & general $V(R,p)$ & -31,049.1 & 2 & -31,030.7\\
\bottomrule
\end{tabular}

  \end{center}
  \begin{tablenotes}[flushleft]
    \footnotesize
   \item \textit{Notes:} Following \citet{chiappori2019aggregate}, the table compares flexible specifications of the utility function \(V(R,p,\vartheta)\), where \(R\) is the return, \(p\) is the winning probability, and \(\vartheta\) is bettor type. The utility function is represented using basis-function expansions, and the table reports log-likelihood values and BIC-based selection for the data used in the two-period betting-model analysis. For each utility specification, \(L=\prod_j \hat p_{w_j j}\), where \(w_j\) is the realized winner in race \(j\) and \(\hat p_{w_j j}\) is the winning probability recovered from the observed odds under that specification. The reported BIC is \(2\log L-k\log N\), where \(k\) is the number of basis-function terms and \(N\) is the number of races; larger values indicate a preferred specification. Expected utility (EU) imposes linearity in \(p\): the homogeneous EU specification uses \(p\,u(R)\), while the heterogeneous EU specification allows \(p\,u(R,\vartheta)\). The homogeneous rank-dependent expected utility (RDEU) specification allows probability weighting through \(G(p)u(R)\), with Yaari's dual theory as the special case \(u(R)=1+R\). The general homogeneous non-expected-utility class does not impose separability between probability weighting and payoff utility. Heterogeneous specifications allow \(V\) to vary with \(\vartheta\); homogeneous specifications do not. The risk-neutral specification is a zero-parameter benchmark. The estimation procedure, including the basis functions and the BIC-based specification search, is identical to that of \citet{chiappori2019aggregate}, with utility normalized at \(R=6\).
  \end{tablenotes}
\end{table}

\end{document}